\documentclass[12pt]{iopart}

\usepackage{iopams}  
\usepackage{amsfonts}
\usepackage{bbm}
\usepackage{color}
\usepackage{xcolor}
\usepackage{todonotes}

\usepackage{algorithm}
\usepackage{algpseudocode}
\usepackage{graphicx}
\usepackage{cite}
\usepackage{ulem}
\usepackage[pdfpagemode=UseNone,
		bookmarksopen=false,
		colorlinks=true,
		urlcolor=blue,
		citecolor=blue,  
		citebordercolor=blue,
		linkcolor=blue, 
		urlcolor=cyan, 
		filecolor=red]{hyperref}

\definecolor{Blue}{rgb}{0.00, 0.00, 1.00}
\definecolor{Red}{rgb}{1.00, 0.00, 0.00}

\begin{document}

\title[Using rare event algorithms to understand extreme heatwave seasons]{Using rare event algorithms to understand the statistics and dynamics of extreme heatwave seasons in South Asia}

\author{Cl{\'e}ment Le Priol$^{1,2}$, Joy M. Monteiro$^{3,4}$, Freddy Bouchet$^2$}

\address{$^1$ Laboratoire de physique à l'ENS de Lyon, UMR 5672, Lyon, France}
\address{$^2$ Laboratoire de Météorologie Dynamique, IPSL, ENS-PSL, CNRS, UMR 8539, Paris, France}
\address{$^3$ Department of Earth and Climate Science, Indian Institute of Science Education and Research Pune, Pune, India}
\address{$^4$ Department of Data Science, Indian Institute of Science Education and Research Pune, Pune, India}
\ead{clement.le-priol@lmd.ipsl.fr, freddy.bouchet@cnrs.fr}

\begin{abstract}

Computing the return times of extreme events and assessing the impact of climate change on such return times is fundamental to extreme event attribution studies. However, the rarity of such events in the observational record makes this task a challenging one, even more so for "record-shattering" events that have not been previously observed at all. While climate models could be used to simulate such extremely rare events, such an approach entails a huge computational cost: gathering robust statistics for events with return time of centuries would require a few thousand years of simulation.

 In this study, we use an innovative tool, rare event algorithm, that allows us to sample numerous extremely rare events at a much lower cost than direct simulations. We employ the algorithm to sample extreme heatwave seasons, corresponding to large anomalies of the seasonal average temperature, in a heatwave hotspot of South Asia using the global climate model Plasim. We show that the algorithm estimates the return levels of extremely rare events with much greater precision than traditional statistical fits. It also enables the computation of various composite statistics, whose accuracy is demonstrated through comparison with a very long control run. In particular, our results reveal that extreme heatwave seasons are associated with an anticyclonic anomaly embedded within a large-scale hemispheric quasi-stationary wave-pattern. Additionally, the algorithm accurately represents the intensity-duration-frequency statistics of sub-seasonal heatwaves, offering insights into both seasonal and sub-seasonal aspects of extreme heatwave seasons. This innovative approach could be used in extreme event attribution studies to better constrain the changes in an event's probability and intensity with global warming, particularly for events with return times spanning centuries or millennia.

\end{abstract}

%
%
\submitto{Environmental Research: Climate}
%
%
%

\section{Introduction}\label{sec:Introduction}

Due to global warming, most regions in the world are experiencing an increase in the frequency and intensity of heatwaves \cite{seneviratne2021} 
with dramatic consequences on human health \cite{romanello2021, dimitrova2021}, crop yields \cite{wegren2011} and ecosystems \cite{ciais2005}. 
Therefore, understanding changes in the probability of occurrence and characteristics of heatwaves is crucial to climate adaptation efforts globally.
While the mechanisms leading to heatwaves are well understood,  research remains to be done to assess at the regional scale the relative importance of each mechanism and how they will be affected by global warming \cite{perkins2015, horton2016}.

South Asia is particularly vulnerable to the impacts of extreme heat. This vulnerability is not just because of the magnitude of heatwaves experienced in this region but also because of its high population density which results in many people being exposed to extreme heat \cite{romanello2021}.

In South Asia, while there does not seem to be a significant increase in exposure to extreme heat (as measured by air temperature alone) over the historical record \cite{rogers2021}, climate models project that exposure to extreme heat will increase under all emission scenarios \cite{mora2017,im2017,mishra2017}. However, few studies focused on the understanding of their mechanisms and of the relative importance of their main precursors \cite{ratnam2016,rohini2016,monteiro2019}.

Complementary to traditional research into the meteorology of extreme events, the field of extreme event attribution (EEA) focuses on assessing the influence of anthropogenic climate change on individual extreme weather events. While the way in which event attribution is framed is debated \cite{hulme2020}, it is generally accepted that
 understanding the changes in probability and intensity of the events and associated physical processes at the regional scale are important both from a scientific and societal perspective.
 
Extreme event attribution research comes in different flavours. Most studies focus on estimating the changes in probability and/or intensity of a given event in the present world with respect to a "counterfactual" world where the global temperature would have remained to its preindustrial value \cite{stott2004,philip2020, zachariah2023a}.
This is known as the \textit{probability-based} approach.
Other studies follow the \textit{storyline} approach which focus on the causal chain leading to the extreme event and how climate change has affected the drivers, like sea surface temperature (SST) or soil moisture deficit for heatwaves \cite{hoerling2013,trenberth2015,shepherd2016}.
Finally, some works condition on the atmospheric circulation associated with the event and evaluate how climate change affected temperature, precipitation and wind speeds conditioned on this circulation \cite{jezequel2018,faranda2022}. 
They also provide information about how rare such atmospheric states are \cite{faranda2023}.

In the probability-based approach, the estimation of unconditional probabilities, or return levels, usually relies on extreme value theory (EVT).
The underlying assumption is that the ground truth distribution of the events lies in the basin of attraction of one of the generalised extreme value distributions \cite{coles2001}.
EVT allows to perform statistical extrapolations to compute the probabilities of unseen events. 
However, it does not actually sample them.
A second limitation of EVT is that the probabilities of unseen, record-shattering, events can be extremely difficult to estimate. For instance, the temperature of the record-shattering 2021 Pacific Northwest heatwave falls outside the range of possible values that could be estimated from the previous years, implying huge uncertainties on estimates of that event's return time \cite{philip2022}.
The possibility of such an event could be anticipated using climate models but without knowledge of its probability \cite{fischer2021}.
\cite{noyelle2023} also showed that the maximal reachable temperature in western Europe are, in many locations, higher than the upper bounds provided by EVT fits. 
As record-shattering events are among the most detrimental \cite{robine2008}, they should be the topic of active research.
In this work, we introduce a new tool, namely \textit{rare event algorithms}, that allows to actually sample extreme events in climate models and provides an unbiased estimator of their probabilities with narrow confidence intervals on the return levels up to return times of millennia or even more.
We argue that this tool could complement EVT in extreme event attribution, especially regarding the most extreme events, the ones with return times of centuries or millennia.

\subsection{Rare event algorithms}

In many fields, the return times for events of interest are orders of magnitude larger than the typical time scale of the dynamics. This is the case, for example, in chemical reactions or the conformal changes of polymers. 
In recent decades, rare event algorithms have been developed to simulate these rare events from the dynamics but at a much lower computational cost than required by direct simulations.
The general principle is to bias an ensemble simulation statistics towards the event of interest without biasing individual simulation members.
For an out-of-equilibrium system such as the climate system, this must be done in the trajectory\footnote{By 'trajectory' we mean the climate model output over time, which is a solution to the dynamical system represented by the climate model.} 
space\footnote{
For an equilibrium system, one knows the probability measure over the state of the system. It is the Gibbs measure: $\mathbb{P}(x) = \exp(\beta H(x)) / Z$ where $x$ denotes the state of the system, $H$ is the Hamiltonian of the system (so $H(x)$ is the energy of the system in state $x$) and $Z$ is a normalisation constant known as the partition function. 
Since one knows the probability measure, one can directly bias the probability measure to sample rare events.
Out-of-equilibrium systems, on the other hand, don’t have a well-defined energy and we don’t know the probability measure. For the climate, the only thing we are able to do is to simulate the dynamics. So we have no choice but to work in the trajectory space.}.
Such algorithms have recently been applied to climate models to sample extreme hot seasons \cite{ragone2018, ragone2020, ragone2021}, extreme winter precipitation \cite{wouters2023} and extreme tropical cyclones \cite{plotkin2019, webber2019}.
Since these algorithms provide sample of actual solutions to the model dynamics, they allow computation of composite statistics on extreme events. Such statistics could, for instance, be used to diagnose how global warming affects different drivers of heatwaves, like the large-scale atmospheric circulation, depleted soil moisture, or SST.

An empirical rare event algorithm based on similar ideas has been proposed to simulate extreme hot seasons from observations and reanalysis data \cite{yiou2020}.
We also note that a different resampling protocol, called ensemble boosting, has been proposed to increase the sampling of extreme events in climate models \cite{gessner2021}, but this protocol does not provide information about the probabilities of the simulated events.

\subsection{Our contribution}
We use the Giardina-Kurchan-Lecomte-Tailleur (GKLT) algorithm \cite{giardina2011, ragone2018, ragone2020, ragone2021, wouters2023} 
to sample extreme heatwave seasons in a subregion of South Asia in the global climate model (GCM) Plasim. 
This algorithm belongs to the family of genetic selection algorithms.
We demonstrate the precision and quality of return time curves and composite maps for heatwaves with 100 year return times that were computed via the algorithm by comparing them with the outcome of a very long (8,000 year) control run. 
For the first time, we compare the return time curve estimate of the algorithm with the one of an EVT fit. We show that the algorithm estimate is both closer to the long control run than the EVT fit and has a much narrower confidence interval.

We show that the algorithm can help to gain insight into the atmospheric circulation associated with extreme heatwave seasons in South Asia. Composite maps created using samples obtained from the algorithm show that extreme heatwaves seasons are associated with an anticyclonic anomaly which is embedded in a large-scale hemispheric quasi-stationary wave-pattern.
We also show that the algorithm correctly represents the intensity-duration-frequency statistics of individual subseasonal heatwaves. Thus, not only does the algorithm capture seasonal statistics accurately, but the results can also be used to estimate the distribution of intensity and duration of heatwaves that comprise an extreme heatwave season.

We define the region and event of interest, and describe the model and data used in section \ref{sec:Event definition, model and data}.  
Section \ref{sec:REA explanation} explains how the rare event algorithm works.
Our results are presented in section \ref{sec:Results}. 
Finally, we discuss the relevance of rare event algorithms for extreme event attribution in section \ref{sec:relevance of EEA} and conclude in section \ref{sec:Conclusion}. A brief explanation of the algorithm is presented in \ref{sec:Details of the algorithm}.

\section{Event definition, model and data}\label{sec:Event definition, model and data}

\subsection{Event definition}\label{sec:event definition}
In South Asia, heatwaves occur during the pre-monsoon season, i.e. March to June, with May-June usually being the hottest months. 
\cite{ratnam2016} identified two distinct heatwave hotspots in India based on the leading EOFs of Tmax variability: one in the northwest and one on the east coast. 
These two regions are also the ones with the highest incidence of mortality attributed to heat waves (see fig. S9 in \cite{justine2023}).
Heatwaves on the east coast of India are affected by the Indian and Pacific oceans and in particular by ENSO. These influences are not represented by our model's setting, which has no dynamical ocean. Due to this limitation, we restrict our study to the northwest region.  

Some heatwave seasons experience multiple heatwaves, whose impact is observed in the seasonal mean as well. For instance, in the pre-monsoon season of 2022, our region of interest (defined below) experienced an unusually warm season with six extreme heat spells \cite{dash2022}. The March-April average temperature was estimated to be a 100-year event in present-day climate \cite{zachariah2023a}. While this is an illustrative example, it underscores the importance of looking not just at individual extreme events but also the extremes of long-term averages. 
 
\begin{figure}[tbh]
\centering
\includegraphics[width=8.3cm]{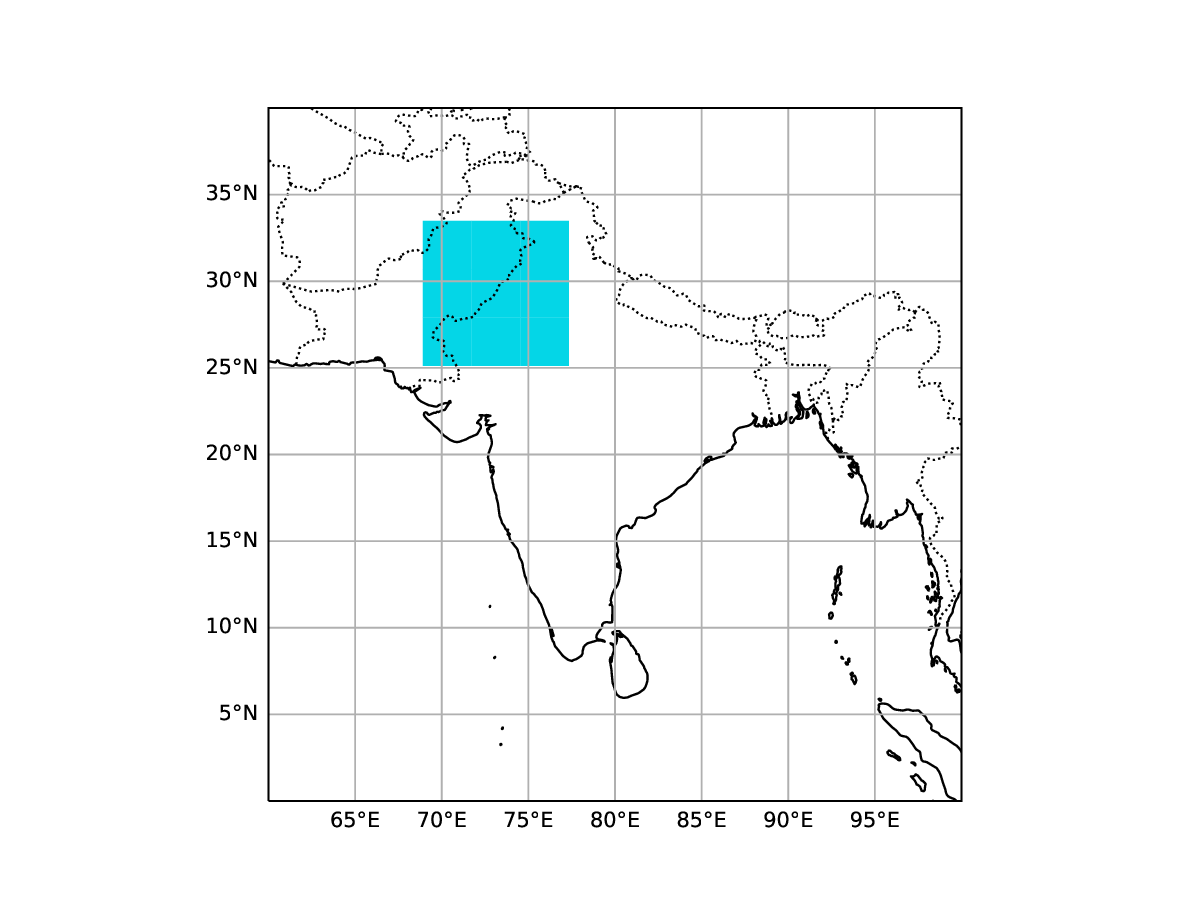}
\caption{The region studied is marked in light blue. \label{fig: region NW India}}
\end{figure}

Our region of interest is defined by a box 70$^\circ$-76$^\circ$E, 25-34$^\circ$N, which includes parts of the Indian states of Rajasthan and Punjab and the Pakistani province of Punjab.
The region is marked in light blue in Fig.~\ref{fig: region NW India}.
Our target observable is the April-May-June (AMJ) average of the area-averaged 2-meter air temperature:
\begin{equation}\label{eq:target observable}
a = \frac{1}{T} \int_{t_0}^{t_0+T} A\left(t \right) \textrm{d}t \quad \textrm{where} \quad 
A\left( t \right)=\frac{1}{\mathcal{A}}\int_{\mathcal{A}}T_{\textrm{2m}} (\textbf{r}, t) \textbf{dr}
\end{equation}
where \textbf{r} is the space vector, $t$ is time, $T_{\textrm{2m}}$ is the 3-hourly 2-meter air temperature, 
$\mathcal{A}$ is the area of the region considered (70$^\circ$-76$^\circ$E, 25-34$^\circ$N), $t_0$ is the 1st of April and $T$ is the AMJ duration, equal to 90 days in the model's setting.

\subsection{Model: Plasim}\label{sec:Model}
All simulations are performed with Plasim \cite{fraedrich2005}, an intermediate complexity climate model developed at the Theoretical Meteorology group of the University of Hamburg. 
Plasim solves the primitive equations for vorticity, divergence, temperature and logarithm of surface pressure. 
The subgrid-scale parameterization is simpler than the ones used by top-class GCM's, allowing it to run at dramatically lower computational cost while preserving the correct large-scale dynamics.
Plasim also features land, ocean, and sea ice components.
We run the model in a stationary climate representative of the late 20th century. The evolution of sea surface temperature (SST) and sea ice follow fixed prescribed seasonal cycles and the CO$_2$ concentration is fixed at 360ppm.
We set the model at T42 horizontal resolution (about 2.8$^\circ$) with 10 vertical sigma-levels. While coarse, this resolution is sufficient to capture large-scale atmospheric dynamics.
The extremely low computational cost of this configuration (less than 1hCPU.yr$^{-1}$) makes it well suited for proof of concept and methodological studies.
Plasim has been used in several studies of large scale phenomena, e.g. to assess wind speed reductions by large-scale wind turbine deployment \cite{miller2016} or to study teleconnections and their changes in a changing climate \cite{herein2017}.
It also proved to be relevant to study large-scale circulations  associated with heatwaves:
the mode 3 quasistationary Rossby wave pattern associated to European heatwaves observed with Plasim in \cite{ragone2018} was confirmed in subsequent studies using CESM \cite{ragone2021} and reanalysis \cite{miloshevich2023a}.
We acknowledge that the coarse resolution and simplified parameterization limit the ability of the model to capture land-atmosphere and boundary layer physics accurately enough to provide quantitatively accurate estimates of extreme heatwave seasons statistics. 
However, the choice of this low computational cost model allows us to demonstrate the efficiency and reliability of the rare event algorithm by comparing its results with an extremely long control simulation.

We produced a control run of 2,000 years (10 independent simulations of 200 years) to serve as initial conditions for the rare event algorithm. 
We also use 1,500 years from this run in fig. \ref{fig:Zg500 100yrs HW  composites} to compare the significance of composite maps obtained using either the rare event algorithm or a control simulation for the same computational cost.
A longer and independent control run of 8,000 years (80 independent simulations of 100 years, \cite{miloshevich2023}) is used as a baseline against which the results of the algorithm are
compared.

\subsection{ERA5 data and detrending}\label{sec:ERA5 data}
\begin{figure}[t]
\centering
\includegraphics[width=6cm]{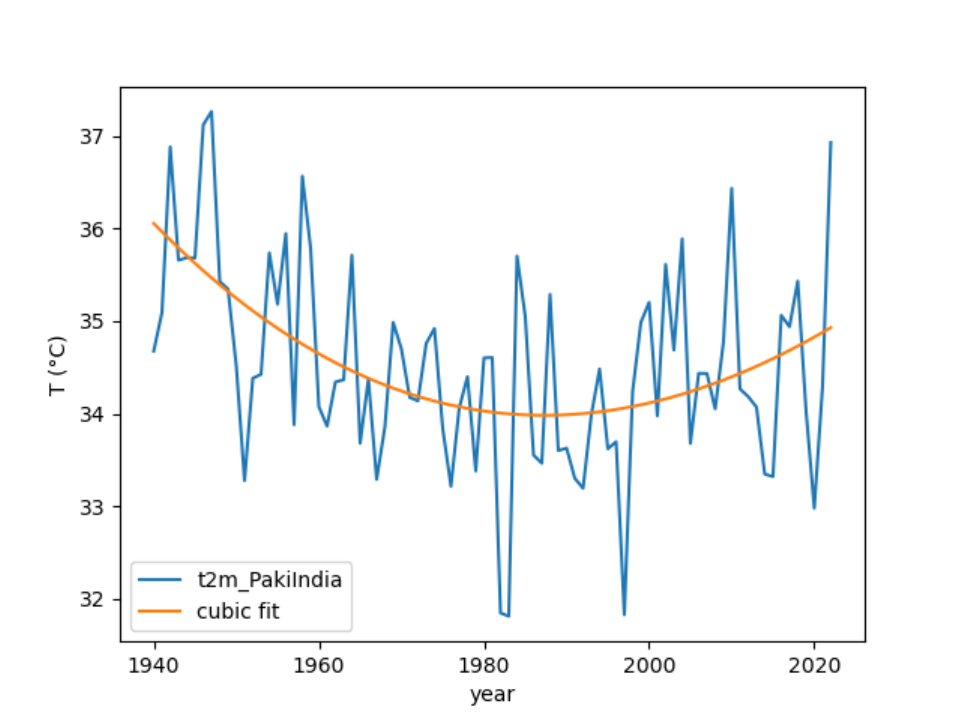} \includegraphics[width=6cm]{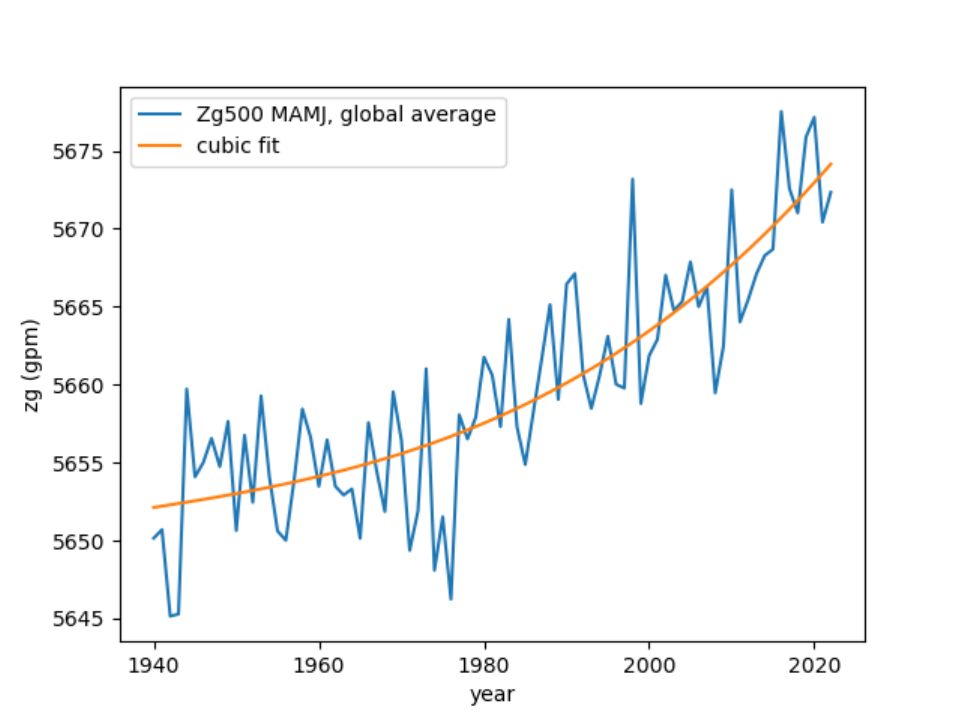}
\caption{Time series of yearly MAMJ average of T2m averaged over the target region \textbf{(a)} and global average of Zg500 \textbf{(b)}. A trend has been computed with a cubic fit (orange lines) and removed from the data before computing the correlation map shown in figure \ref{fig:Zg500 correlations maps Plasim ERA5}. \label{fig:historical trends}}
\end{figure}

We use ERA5 reanalysis dataset \cite{hersbach2020} to assess the ability of Plasim to represent teleconnection between the surface air temperature over the target region and the geopotential height.
We use daily 2-meter air temperature (T2m) and geopotential at 500hPa. 
The temperature is selected at 9h UTC (14h30 Indian time)  which corresponds to the approximate time of peak temperature over the target region.
The geopotential, which does not have a daily cycle, is selected at 12h UTC.
The horizontal resolution is $0.25^{\circ}\times0.25^{\circ}$.
We detrended the data by performing a cubic fit to the time series of yearly MAMJ averages of T2m averaged over the target region and of the global average of Zg500 (see fig. \ref{fig:historical trends}). The resulting trends were then subtracted uniformly to all days and grid points.
Our region of interest appears to partly have a positive and part negative long-term temperature trend \cite{ross2018}. These competing trends are due to aerosols and may be responsible for the nonlinear trend that we observe. The geopotential trend is due to the thermal expansion of the lower atmosphere caused by global warming and is seen in multiple reanalysis dataset for both annual and seasonal means \cite{christidis2015}.

We have added the month of March to this comparison in order to have more points and to increase the significance of the ERA5 map. Although March is not included in our event definition, it also belongs to the heatwave season in South Asia. Therefore, it makes sense to include it in our comparison as we want to assess Plasim's ability to represent the teleconnection during the heatwave season.

\section{Rare event algorithm}\label{sec:REA explanation}

We use the Giardina-Kurchan-Lecomte-Tailleur (GKLT) algorithm \cite{giardina2011}.
It is a genealogical selection algorithm, adapted from \cite{delmoral2005} by \cite{giardina2011} to sample rare extrema of large-time averages of a target observable $A(t)$. 
We explain below how the algorithm works.
The term \textit{trajectory} denotes a model output over time, which is a solution to the dynamical system represented by the climate model

Consider first a classical ensemble simulation.
Most trajectories will produce typical events and only a handful will yield extreme events. 
For instance, if we are interested in a centennial event,
only one out of hundred trajectories will sample the event of interest and 99\% of the trajectories will be wasted data and computation time.
To save computation time, it is more efficient to discard, early in the simulation, trajectories that will not yield the event of interest and to focus instead on the ones expected to yield the desired event.
This is precisely what a genealogical selection algorithm aims at doing.

\begin{figure}[tbh]
\centering
\includegraphics[width=12cm]{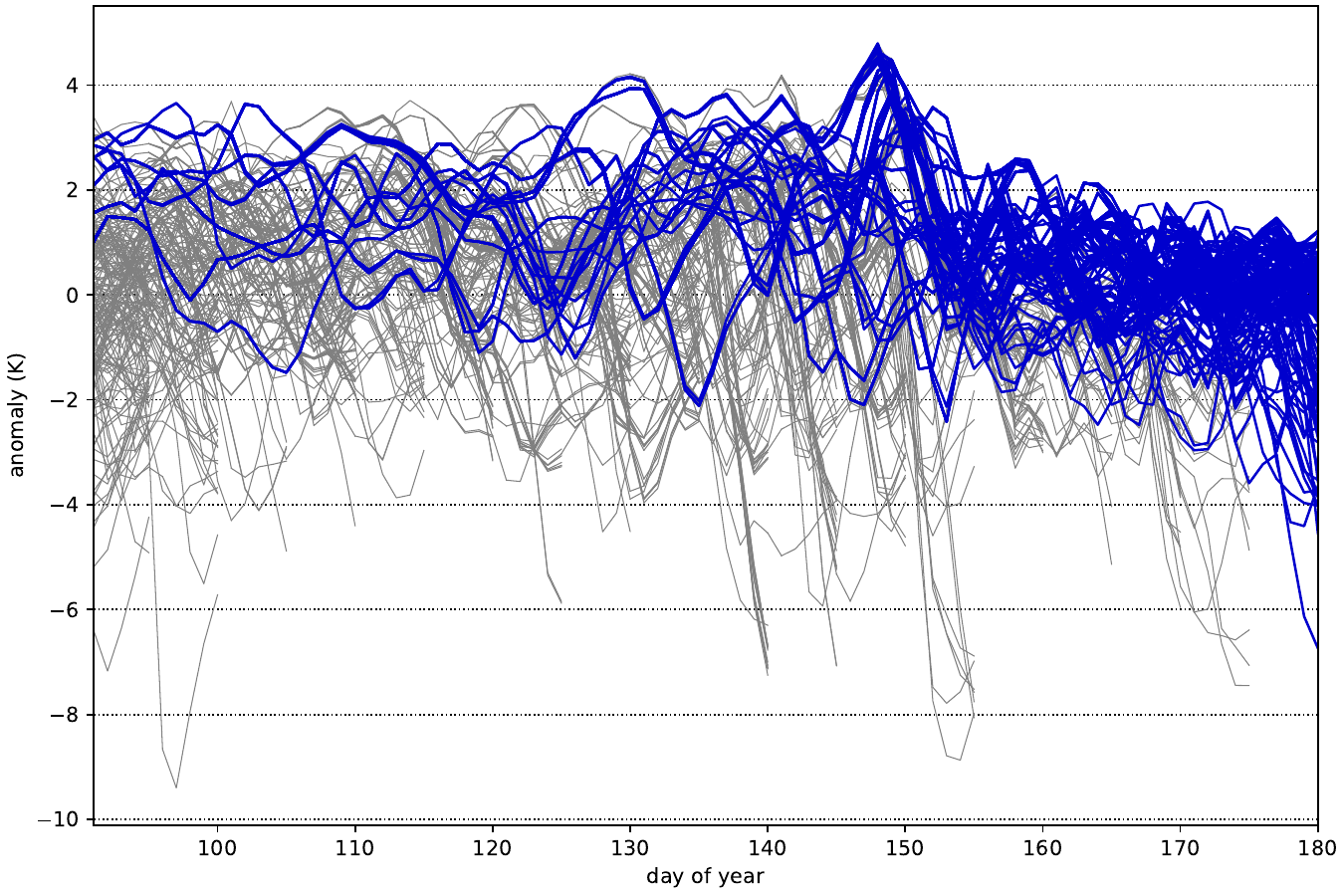}
\caption{ 
Time series of anomaly of daily averaged surface air temperature for one realisation of the rare event algorithm.
$N=200$ trajectories are initiated from $N$ distinct years from the control run.
Resampling is performed every 5 days: some trajectories are discarded while others are duplicated. 
Trajectories that survive until the end of the algorithm are highlighted in blue. Grey lines correspond to trajectories that are eventually discarded. 
The number of trajectories remains constant throughout the simulation.
\label{fig:algorithm illustration}}
\end{figure}

Figure \ref{fig:algorithm illustration} illustrates the behaviour of the algorithm. 
$N=200$ trajectories are simulated in parallel. Every $\tau=5$ days, some trajectories are discarded while others are duplicated in such a way that the number of trajectories remains constant.
Most trajectories are discarded at some point before the end of the algorithm while a handful of initial members are duplicated at every step and survive until the end of the algorithm. 

To be more specific, let $x(t)$ be a vector that represent the state of the model at time $t$ (in practice, $x(t)$ is the vector of values of all prognostic variables used to integrate the model). 
For any time-dependent observable $A$, we may make implicit the dependence on the system state by writing $A(t)$ instead of $A\left ( x(t) \right)$.
We simulate $N$ trajectories in parallel, starting from $N$ independent initial conditions. 
Trajectories are integrated in pieces of duration $\tau$ for a total time $T_a  =I\tau$.
$\tau$ is called the \textit{resampling time} of the algorithm.
At each resampling time step $t_i=i\tau$, we assign to the $n^{th}$ trajectory a weight
\begin{equation}\label{eq:weights formula}
    W_n^{(i)} = \frac{1}{Z_i}\exp \left( k \int_{t_{i-1}}^{t_i}  A\left(x_n(t) \right) dt \right)
\end{equation}
where $Z_i$ is a normalization factor such that $\sum_{n=1}^N W_n^{(i)} = N$.
Each trajectory is then assigned a random number of clones $m_n^{(i)}$ such that 
$\mathbb{E}[m_n^{(i)}] = W_n^{(i)}$ and $\sum_{n=1}^N m_n^{(i)} = N$.
Trajectories with small weight ($W_n^{(i)} \ll 1$) are likely to be discarded ($m_n^{(i)}=0$) while trajectories with large weights are duplicated.
The \textit{biasing parameter} $k$ controls the strength of the selection.
The larger $k$, the more stringent is the selection of trajectories that are duplicated, leading to a larger probability for trajectories with very large time-average of $A(t)$.
Since the dynamics of the climate model is deterministic, we need to add a small perturbation after the resampling step in order to allow clone trajectories to separate in the next integration step\footnote{The perturbation is introduced by adding to the coefficients of the spherical harmonics of the
logarithm of the surface pressure a set of random numbers, sampled independently according to a
uniform distribution in $ \left[-\epsilon \sqrt{2}, \epsilon \sqrt{2} \right] $,
with $\epsilon = 10^{-4}$ following the implementation of \cite{ragone2018}.
Note that the perturbation method must be adapted to each model.}.
The algorithm samples extrema of $\frac{1}{T_a} \int_{t_0}^{t_0+T}  A(t) dt$ by 
favouring the selection of trajectories with persistent large values of $A(t)$.
A pseudo-code of the algorithm is given in algorithm \ref{GKLT algorithm}.

\begin{algorithm}
\caption{GKLT algorithm}
\label{GKLT algorithm}
\begin{algorithmic}
\State Initiate $N$ trajectories from $N$ independent initial conditions: $x_n (t=0)=x_n^0$ for $1 \leq n \leq N$.
\For {$i=1$ to $I$} 
	\State (i) Integrate the trajectories between $t_{i-1}=(i-1)\tau$ and $t_{i}=i\tau$.
	\State (ii) Assign to each trajectory a weight $\tilde{W}_n^{(i)} = \exp \left( k \int_{t_{i-1}}^{t_i}  A\left(x_n(t) \right) dt \right)$
	\State (iii) Normalize the weights: $W_n^{(i)}=\tilde{W}_n^{(i)}/Z_i$ .
	where 	$Z_i= \frac{1}{N} \sum_{n=1}^N \tilde{W}_n^{(i)}$ .
	\State (iv) For each trajectory draw a number of clones $m_n^{(i)} = \lfloor W_n^{(i)} + u \rfloor$ where $u \sim \mathcal{U}([0,1])$. 
	\State (v) Duplicate or eliminate clones to keep a constant number of trajectories:
	Compute $\Delta N_i = \sum_{n=1}^N m_n^{(i)} - N$ ; if $\Delta N_i > 0$, select $\Delta N_i$ clones without repetition and delete them ; if $\Delta N_i < 0$, select $-\Delta N_i$ clones with repetition and duplicate them.
    \State (vi) Add a small random perturbation to each clone so that copies of the same trajectory separate when we continue the model integration.
\EndFor \\
\textbf{Note:} Trajectories for which $m_n^{(i)}=0$ are discarded and replaced by copies of trajectories for which $m_n^{(i)} \geq 2$.
\end{algorithmic}
\end{algorithm}

Note that we know the probabilities with which we discard or duplicate trajectories at each resampling step. This knowledge allows us to compute a probability for each trajectory surviving until the end of the algorithm, which in turn enables us to compute various statistics such as return times or any composite statistics (see \ref{sec:Details of the algorithm} for details about the computation of the probabilities).
This is where the added value of rare event algorithms lies, in comparison to other approaches.

Our target observable $A(t)$ is the surface air temperature averaged over the domain 70°-76°E, 25-34°N,
(see figure \ref{fig: region NW India}).
We performed $M=6$ rare event experiments with $N=200$ trajectories for a total time $T_a=90$ days from April 1st to June 30th\footnote{The number of days per month in the simulation is constant, equal to 30.}. 
The resampling time is $\tau=5$ days which corresponds to the synoptic time scale of the atmospheric dynamics.
The biasing parameter is $k=0.1$ K$^{-1}$.day$^{-1}$. 
The parameter values are summarised in table \ref{tab:parameter values}.
For each experiment $m=1..M$, the initial condition of the n$^{th}$ trajectory is taken as the 1st of April of the year $n$ of control run $m$. 
Since there is no oceanic variability in our simulations, the longest time scale is the one associated to soil moisture variability which is seasonal. 
Therefore we can consider that the 6 experiments start from 6 independent sets of 200 independent initial conditions. 
The computational cost of each experiment corresponds to the simulation of 200 AMJ seasons, equivalent to 50 years of simulation, leading to a total cost of 300 simulated years.

\begin{center}
\begin{table*}
\centering
    \begin{tabular}{| c | c | c | c | c | c | c |}
        \hline
        Parameters & $N$ & $k$ & $\tau$ & $T_a$ & $I$ & $M$ \\ \hline		
    	Values & 200 & 0.1 K$^{-1}$.day$^{-1}$ & 5 days & 90 days & $T_a/\tau = 18$& 6 \\ 
    	\hline
    \end{tabular}
    \caption{Values of the parameters of the rare event algorithm used in this study. $N$ is the number of trajectories, $k$ the biasing parameter, $\tau$ the resampling time, $T_a$ the total integration time, $I$ the number of resampling steps and $M$ the number of independent realisations of the algorithm. \label{tab:parameter values}}
\end{table*}
\end{center}

\paragraph{Possible range of the parameters}
Let us discuss the possible ranges of the parameters introduced above.
The algorithm provides unbiased estimators of rare events probabilities with typical error of order $1/\sqrt{N}$ \cite{delmoral2011}, so the larger $N$, the more accurate the estimates are.
In practice $N$ should be at least a hundred and could typically range up to 1000 trajectories for a climate model. The feasible value depends on the model numerical cost and computation limitations.
The available initial conditions may be split into $M$ independent realisation of the algorithm to facilitate the computation of confidence intervals on the algorithm's results. So $M= N_{IC}/ N$ where $N_{IC}$ is the number of independent initial conditions available and $N$ is the number of trajectories in each realisation of the algorithm.
The resampling time $\tau$ must be large enough so that trajectories have time to separate between successive resampling steps. A typical value for atmospheric dynamic is the synoptic time scale, i.e. $\tau$ should be of the order of a few days.
In order to sample extremes of large time averages of $A(t)$, there must be a reasonable number $I$ of selection steps.
In practice, we found that $I$ should be at least 10 (see section \ref{sec:Methodological study}). This sets a lower bound constraint on $T_a = I \tau$. There is no upper bound other than computation limitations. For instance, one could set $T_a=365$ days to sample anomalously hot years.
Finally, the value of the biasing parameter $k$ must be adapted to the targeted event. A rough estimate is that $k$ must be of the order of $a/(T_a \Sigma^2)$ where $\Sigma$ is the standard deviation of $\frac{1}{T_a} \int_{t_0}^{t_0+T}  A(t) dt$ and $a$ is the targeted temperature anomaly (see \ref{sec:Details of the algorithm} for a derivation of this estimate). However, note that as $k$ increases, the weights are concentrated on a smaller number of trajectories and fewer trajectories survive at each step. So a trade-off must be found between the values of $k$ and $N$ to avoid that all surviving trajectories evolve from a single initial condition.

\section{Results}\label{sec:Results}

\cite{ragone2018,ragone2021} have assessed the reliability of the rare event algorithm for European extreme warm summers
by comparing its results against a long control run of 1,000 years. 
Here we demonstrate the relevance of the algorithm for South Asian extreme heatwave seasons. 
We use a longer available control run of  8,000 years  to extend the comparison to larger return times.
We compare the return time curve estimated from the algorithm with an EVT fit and show that the algorithm estimate is more reliable with a much narrower confidence interval.
The long control run also provides robust estimates of centennial heatwave season composite maps against which we can compare the composite maps derived from trajectories selected by the algorithm.
Finally, we show that the algorithm also correctly estimates the intensity-duration-frequency statistics of subseasonal heatwaves.

We refer to appendix \ref{sec:Details of the algorithm} or to the Supplementary Information of \cite{ragone2018} for the details of how to compute return time curves and composite statistics from the algorithm.

\subsection{Importance sampling and return time curves}\label{sec:importance sampling & RT curves}

\begin{figure}[tbh]
\centering
\includegraphics[width=0.6\linewidth]{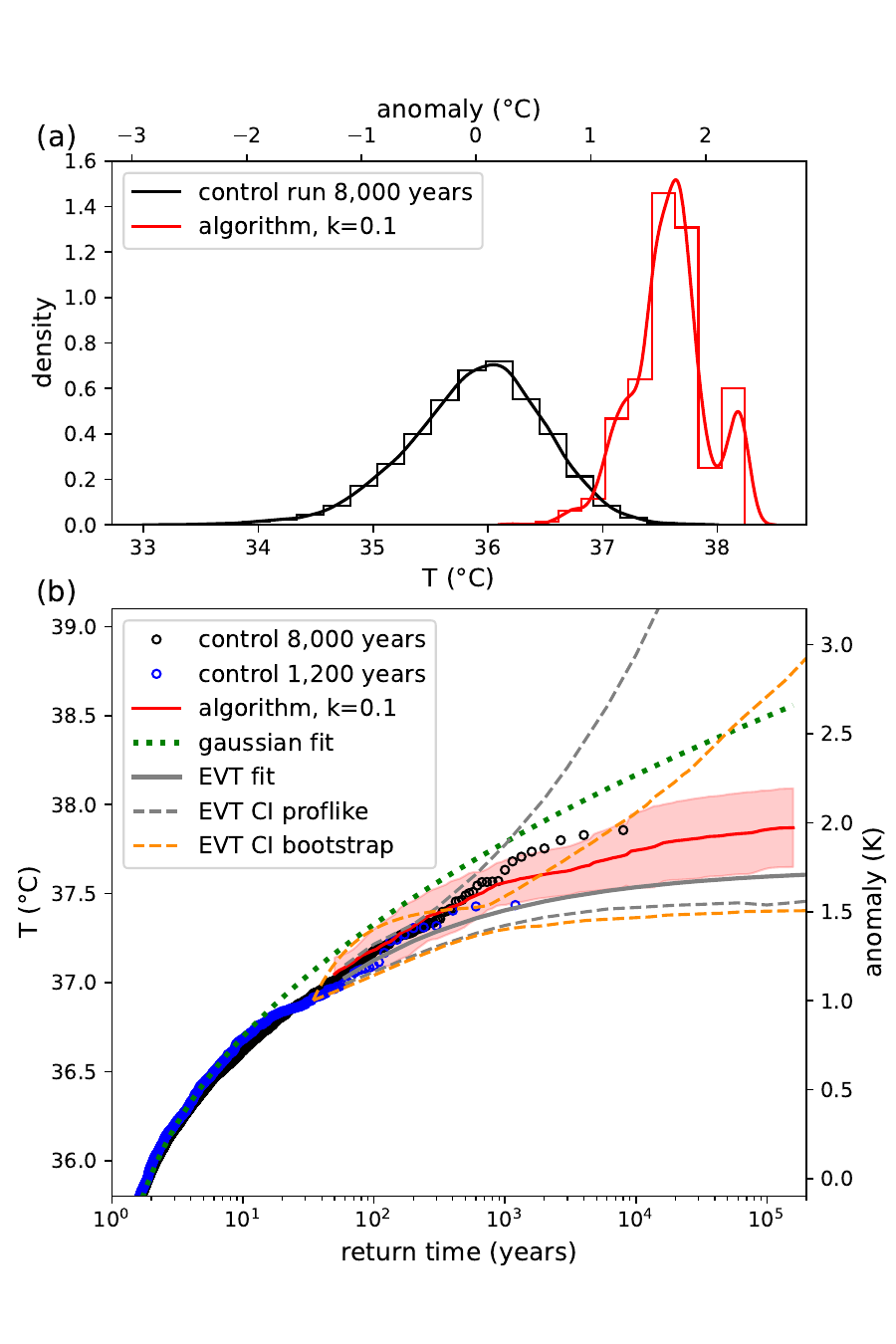}
\caption{\textbf{(a)} Histograms and kernel density estimates of the AMJ averaged surface air temperature $a$ in the 8,000-year control run (black) and the 6 rare event algorithm experiments  with k=0.1 K$^{-1}$.day$^{-1}$ (red). 
\textbf{(b)} Return time curve of $a$ built from the 8,000-year control run (black circles) and an independent 1,200-year control run (blue circles).
The 1,200-year run was used to perform an EVT fit (grey) and served as initial conditions for the 6 algorithm experiments (red). Red shaded area corresponds to the 95\% CI of the algorithm built using Student's t law. 
95\% CI for the EVT fit (dashed lines) is estimated via two different methods: (i) profile likelihood (grey) and (ii) a non-parametric bootstrap (orange).
\label{fig:PDF & RT curves}}
\end{figure}

Figure \ref{fig:PDF & RT curves}a) shows the distributions of AMJ averaged temperature for the control run and the rare event algorithm (the 6 experiments have been pooled together). 
The algorithm effectively samples events above 1K anomaly and up to above 2K, a value never reached in the 8,000-year long control simulation.
Most events lie in the far upper tail of the control run distribution ($\sigma_{\textrm{CTRL}} = 0.60$ K.). This represents a considerable gain in computation time for the simulation of these extreme events.

On figure \ref{fig:PDF & RT curves}b), we compare different estimations of the return time curve of AMJ averaged temperature $a$. 
The black and blue circles correspond to the 8,000-year and an independent 1,200-year long control runs respectively. 
This 1,200-year run provided initial conditions for the algorithm whose return time curve (red line) is obtained by averaging together the 6 independent experiments. The spread of the experiments allows to build a 95\% confidence interval (CI) on the return levels (red shaded area). Details about the estimation of the CI are provided in \ref{sec: Appendix confidence and significance}.
The algorithm and control run curves are in excellent agreement in the range 100-1,000 years. 
The two curves separate for return times larger than 1,000 years. However, the control run estimates for these extremely rare events are not very reliable since we are on the far edge of the empirical distribution. 

How good is the rare event algorithm estimate compared to other approach?
To answer this question, we added on the panel the return time curves of two statistical fits performed on the same 1,200-year run that provided initial conditions for the algorithm. 
The first one is a gaussian fit (green dotted line). While a gaussian fit may be a good approximation in the bulk of the distribution, it fails to capture the statistics of the extremes by overestimating their return levels.
The second is an EVT fit (grey line) using the peak over threshold approach with a threshold of 36.9°C. The parameters are estimated by maximum likelihood estimation (MLE). 
The algorithm central estimate is much closer to the 8,000-year run return time curve than the EVT fit.
The grey and orange dashed lines represent the bounds of the EVT fit 95\% confidence interval computed with two different methods: profile likelihood (grey) and a non-parametric bootstrap (orange).
The non-parametric bootstrap is widely used in the climate literature to compute confidence intervals on EVT fits (e.g. \cite{philip2020, noyelle2023}).
The profile-likelihood method consists in finding the range of parameters such that the likelihood exceeds a certain threshold \cite{coles2001} and seems more grounded theoretically.
Details about the computation of the confidence interval and support for the choice of the threshold value are provided in \ref{sec: EVT fit details}.
The bootstrap CI is too narrow for return times of order of millennia, failing to capture the empirical distribution of the 8,000-year run, before widening rapidly for larger return times.
The profile-likelihood CI widens extremely fast for return time larger than a few centuries\footnote{
The upper bound of the CI corresponds to values of $\xi>0$. As the shape parameter is almost always found to be negative in heatwave analyses \cite{vanoldenborgh2022}, one could envision to constrain the result on $\xi \leq 0$. This possibility is not implemented in the package we used. However, an upper bound with $\xi=0$ (exponential decay) would correspond to a straight line in the plot, still a rapidly growing CI.
}
but has the merit of capturing the 8,000-year empirical return time curve.
The algorithm confidence interval mostly captures the empirical distribution.
Note that the empirical return time curve is not the ground truth: there may be fluctuations around the true return time curve, especially for the most extreme points.
Moreover the algorithm CI is much narrower than the EVT ones for return times greater than 600 and $10^4$ years, for the profile-likelihood and bootstrap methods respectively, and widens at a drastically lower pace.
We conclude that the algorithm provides a much more precise estimate of the return time curve than the gaussian and EVT fits. Indeed, the central estimate is in better agreement with the empirical curve while the algorithm's CI is narrower than the one obtained from the EVT fit with the profile-likelihood. The fact that the non-parametric bootstrap confidence interval fails to capture the empirical distribution suggests that this method is not appropriate to estimate confidence intervals.

\subsection{Composite statistics of centennial heatwave seasons}\label{sec:Statistics of centennial heatwave seasons}

\setcounter{footnote}{0} 

As the algorithm produces heatwave seasons which are solutions to the model dynamics, we have access not only to the AMJ averaged temperature but to any observable at the model output frequency.
This allows to compute a variety of composite statistics conditioned on extreme heatwave seasons. 
In this section, we compare composite statistics conditioned on centennial heatwave seasons computed from the rare event algorithm and from the control run.
We define \textit{centennial heatwave seasons} as the seasons for which the AMJ averaged temperature $a$ is higher than a threshold $a_{100}$.
We use a common threshold value which is the empirical 100-year return level of the 8,000 year long control run: 
$a_{100}=37.18$°C corresponding to an anomaly of 1.28 K.

We study the composite statistics of the seasonal evolution of the temperature, of sub-seasonal heatwaves, and of the seasonal average of Zg500.
We find that the statistics of trajectories generated by the algorithm are in very good agreement with the ones from the control run. In particular, the algorithm correctly reproduces the seasonal evolution of the temperature during 100-year heatwave seasons as well as the frequency-duration-intensity statistics of sub-seasonal heatwaves.
Therefore the algorithm can be used, not only to compute the average temperature anomaly of a heatwave season, but also to anticipate what can be the intensity and duration of heatwaves within the season.

\subsubsection{Seasonal evolution of the temperature}\label{sec:trajectories of centennial heatwave seasons}

\begin{figure}[tbh]
\centering
\includegraphics[width=1\linewidth]{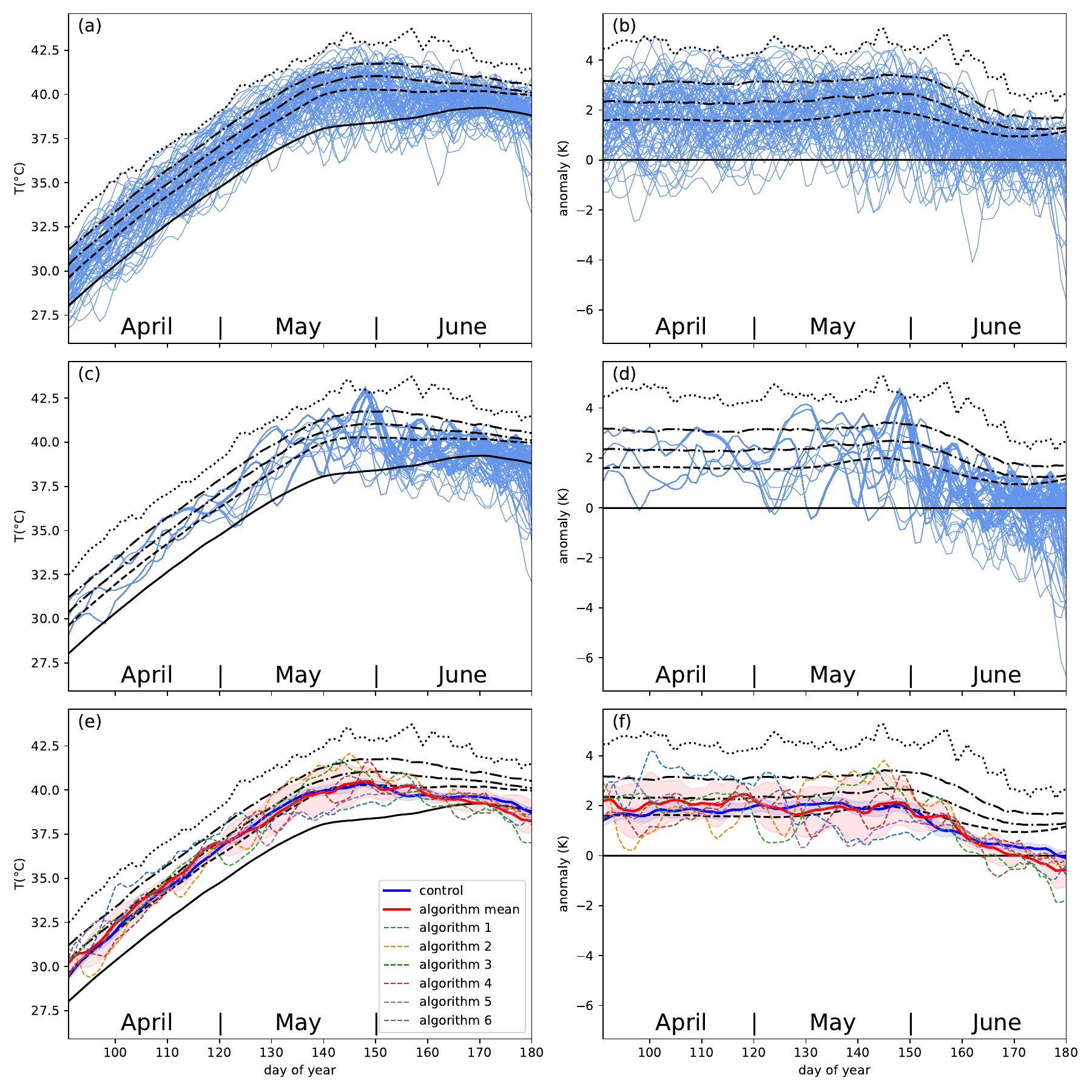}
\caption{Daily-averaged temperature of centennial heatwave seasons in the control run \textbf{(a-b)} and one rare event algorithm experiment \textbf{(c-d)}.
The left column shows the absolute temperature while the right one shows the temperature anomaly with respect to the climatology. Black lines correspond to the climatology (solid), one standard deviation (dashed), the 95th and 99th percentiles (dash-dotted) and the maximum (dotted) of the long control run daily temperature distribution. 
Panels \textbf{(e)} and \textbf{(f)} show the composite mean of daily temperature conditioned on centennial heatwave seasons in the control run (blue) and the algorithm (red) as well as the mean in each individual experiment (dashed coloured lines). The blue and red shaded bands on panel \textbf{(f)} represent the 95\% confidence interval for the control run and the algorithm respectively.
Note that the months of April, May and June correspond respectively to day of year 91-120, 121-150 and 151-180.
\label{fig:traj_CTRL_vs_GKTL}}
\end{figure}

We first compare the seasonal evolution of the temperature for the centennial heatwave seasons from the control run and for the ones from the algorithm in figure \ref{fig:traj_CTRL_vs_GKTL}.
Note that the months of April, May and June correspond respectively to day of year 91-120, 121-150 and 151-180.
Panel (a) shows the daily-averaged absolute temperatures for the 80 centennial heatwave seasons from the control run. Panel (b) shows the corresponding anomalies with respect to the climatology.
The bulk of the trajectories concentrate between the climatology (solid black line) and the 99th percentile (upper dashed-dotted line). 
The temperature anomalies decrease towards the end of the season with more trajectories being below the climatology and almost no one being above the 99th percentile.
Panels (c) and (d) show the same quantities for the centennial heatwave seasons produced by one realisation of the algorithm. 
Out of 200 trajectories, 161 correspond to centennial heatwave seasons. 
At the beginning of the season, the distribution of temperature is more restricted in the algorithm than in the control run because the number of independent trajectories is small (here all centennial heatwave seasons come from 4 independent initial conditions).
The number of separated trajectories increases progressively during the simulation.
All trajectories remain between the climatology and the 99th percentile during most of the time. The temperature anomaly also decline at the end of the season.
However, note that visual comparison can be misguiding to interpret the distribution of trajectories produced by the algorithm. Indeed, trajectories in the control run have equal probability while the ones produced by the algorithm have different probabilities which can differ by orders of magnitude but these differences are not visible on the figure.
In panels (e-f) we make a more quantitative comparison by plotting the composite mean temperature of centennial heatwave seasons, along with a 95\% confidence interval in panel (f). 
The composite means of the control run and the algorithm are close to one another and consistent within the uncertainty range from the algorithm.

We notice that, while the climatological mean of daily-averaged temperature keeps increasing along the season until the end of June, extreme heat (characterized by the 95th and 99th percentiles of daily temperature) peaks end of May and slowly decays afterwards. As a consequence, the variability of temperature is lower end of June.
Regarding centennial heatwave seasons, the conditioned mean temperature is consistently at least one standard deviation higher than the climatology during the months of April and May (day of year 91 to 150). It peaks end of May and then decays, at a faster rate than extreme quantiles, to reach a value close to the climatology by the end of the season. 
This means that the conditions leading to anomalously high temperatures in April-May are different from the ones leading to high temperature anomalies in June.

\subsubsection{Sub-seasonal heatwave statistics}
During a heatwave season, periods of hot or extremely hot temperatures alternate with relatively cooler periods. 
Long heatwaves are the most detrimental to human health. 
Therefore, knowing that we have an extreme heatwave season, 
it is also important to correctly represent the sub-seasonal statistics of heatwaves.
Hereafter, we call heatwave a period during which the daily-averaged temperature remains above a (time-dependent) threshold. We do not impose any restriction on the duration of the heatwave.
We use two different thresholds: $A_{\textrm{th}}(t) = A_{\textrm{mean}}(t) + \sigma_A(t)$ and
$A_{\textrm{th}}(t) = A_{q95}(t)$ where 
$A_{\textrm{mean}}(t)$, $\sigma_A(t)$ and $A_{q95}(t)$ are the climatological mean, standard deviation and 95th percentile of the long control run daily temperature distribution.
We look at events where the temperature exceeds these thresholds.
For the sake of convenience, we call these events \textit{above 1$\sigma$ events} and \textit{above $q^{95}$ events} respectively. 
When we don't need to be specific, we simply call them \textit{heatwaves}.

\begin{figure}[tbh]
\centering
\includegraphics[width=12cm]{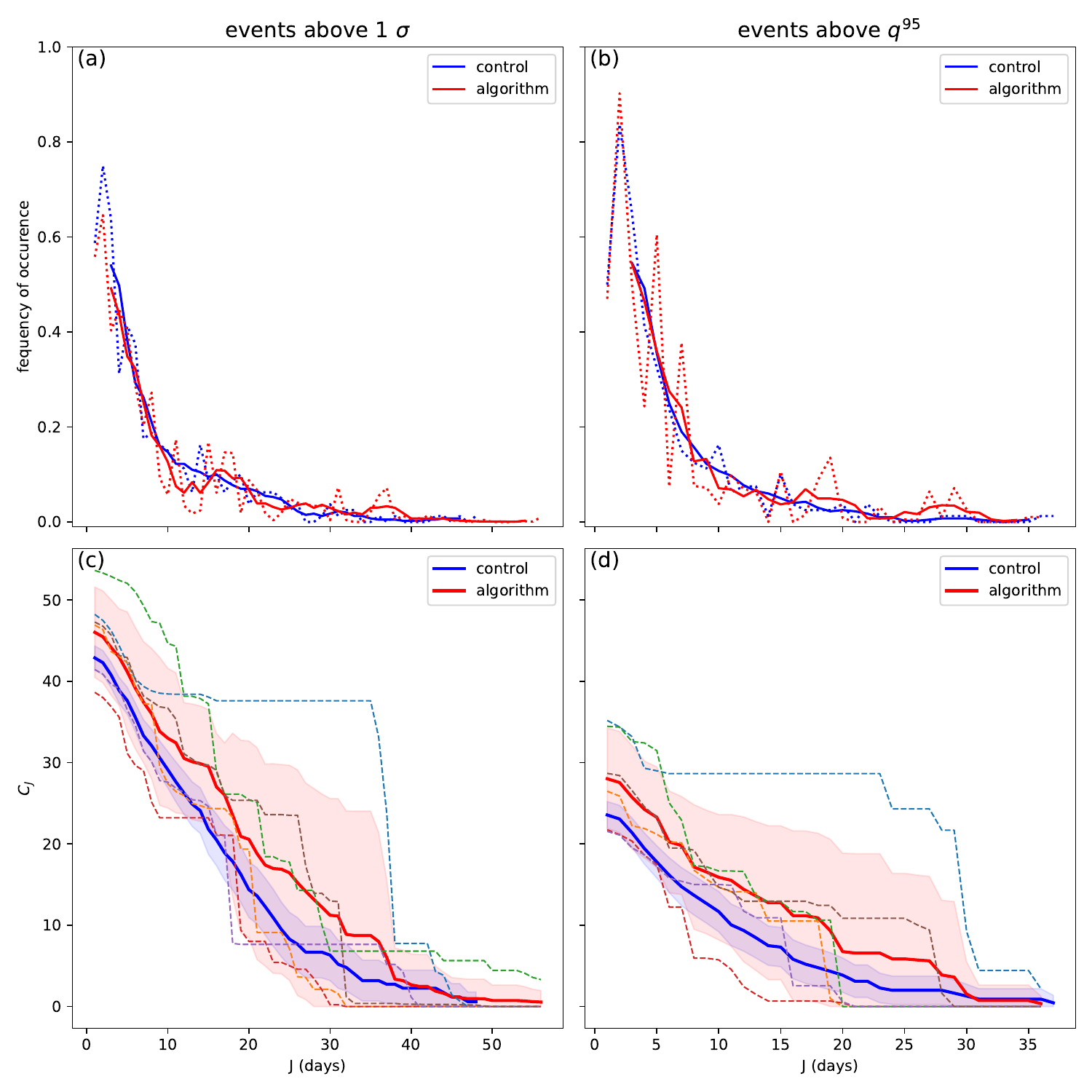}
\caption{
\textit{Top:} Frequency of heatwave occurrence as a function of the heatwave duration $J$. The dotted lines represent the direct estimate computed with equations \ref{eq: frequency estimator}.  The solid lines represent a smoothed version, using a 5-day centred mean.
\textit{Bottom:}  Cumulative number of days $C_J$ belonging to heatwaves lasting $J$ days or more. Dashed lines correspond to the individual realisation of the algorithm. The blue and red shaded area correspond to the 95\% confidence interval from the control run and the algorithm respectively.
\label{fig:subseasonal heatwaves statistics}}
\end{figure}

For $1 \leq J \leq T_a=90$ we define the observable $\# \textrm{HW}_J \left(\{x(t)\}_{0\leq t \leq T_a} \right)$ 
which counts the number of heatwaves lasting exactly $J$ consecutive days along the season $\{x(t)\}_{0\leq t \leq T_a}$.
We first study the average number of heatwave lasting exactly $J$ days during centennial heatwave seasons:
\begin{equation}\label{eq: frequency definition}
f_J := \mathbb{E}\left[ \# \textrm{HW}_J \left(\{x(t)\}_{0\leq t \leq T_a} \right) | a\left(\{x(t)\}_{0\leq t \leq T_a} \right) > a_{100} \right] .
\end{equation}
We compute the empirical estimation of $f_J$ from the control run and from the algorithm using the following formula:
\begin{eqnarray}\label{eq: frequency estimator}
\hat{f}_{J,ctrl} &= \frac{1}{K} \sum_{k=1}^K  \# \textrm{HW}_J \left(\{x_k(t)\}_{0\leq t \leq T_a} \right) \\
\hat{f}_{J,algo} &= \sum_{n|a_n>a_{100}} w_n \# \textrm{HW}_J \left(\{x_n(t)\}_{0\leq t \leq T_a} \right) \quad \textrm{with } w_n = \frac{p_n}{\sum_{n|a_n>a_{100}} p_n} , \\
p_n &= \frac{\exp \left( k\int_0^{T_a} A(t)dt \right)}{N\tilde{Z}} 
     = \frac{\exp \left( kaT_a\right)}{N\tilde{Z}}.
\end{eqnarray}
In the first line, the sum runs over the K=80 years whose average temperature is above the 100 years return level $a_{100}$.
The $p_n$'s are the probabilities attached to the algorithm trajectories and $\tilde{Z}$ is a normalization factor that is different for each realisation of the algorithm.
We refer to \ref{sec:Details of the algorithm} for the details of the probabilities computation.

The top row of figure \ref{fig:subseasonal heatwaves statistics} shows the empirical frequencies $\hat{f}_{J,ctrl} $ and $\hat{f}_{J,algo}$ for the two heatwave thresholds.
The algorithm and the control run are in excellent agreement:
both the number of heatwave days and the decreasing trend are correctly captured.
The second row of figure \ref{fig:subseasonal heatwaves statistics} shows the total number of days that belong to heatwaves that last $J$ days or more:
\begin{equation}
C_J = \sum_{j\geq J} j\times f_j .
\end{equation}
In particular, $C_1$ is the conditional expectation of the total number of heatwave days during centennial heatwave seasons.
The empirical estimate from the algorithm is 4 to 5 days higher than the control run for $J < 37$ (above $1\sigma$ events) or 
$J < 30$ (above $q^{95}$ events). This can be traced to a larger occurrence of heatwaves of duration $J=32, 37$ (above $1\sigma$ events) 
and $J=27, 29$ (above $q^{95}$ events).
The control run curve is nevertheless within the uncertainty range of the algorithm, with 2 out of 6 algorithm realisations yielding estimates of $C_1$ that are lower than the control run value.

\begin{figure}[tbh]
\centering
\includegraphics[width=12cm]{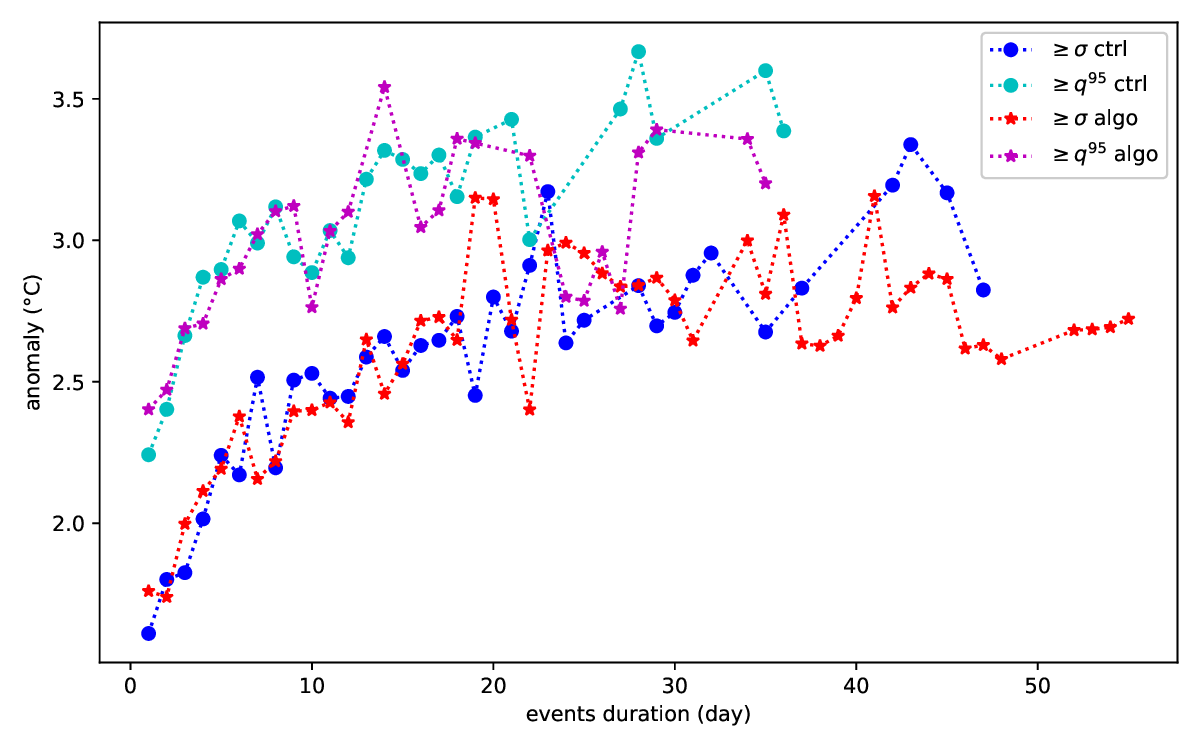}
\caption{\textbf{Heatwaves intensity versus heatwaves duration}. Mean temperature anomaly of heatwaves versus their duration. \label{fig:subseasonal heatwaves intensity}}
\end{figure}

Finally, we assess the ability of the algorithm to correctly represent the intensity-duration relationship.
Figure \ref{fig:subseasonal heatwaves intensity} shows the mean temperature anomaly of heatwaves as a function of their duration.
Once again the algorithm is in excellent agreement with the control run despite the noisy curves (due to the low number of heatwaves for each duration).
In particular we observe a first regime of rapid increase of the intensity with the events duration up to $J=8$ days before a second regime of slower increase.

\subsubsection{Composite maps and hemispheric quasi-stationary wave-pattern}\label{sec:teleconnection patterns}

\begin{figure}[tbh]
\centering
\includegraphics[width=0.49\linewidth]{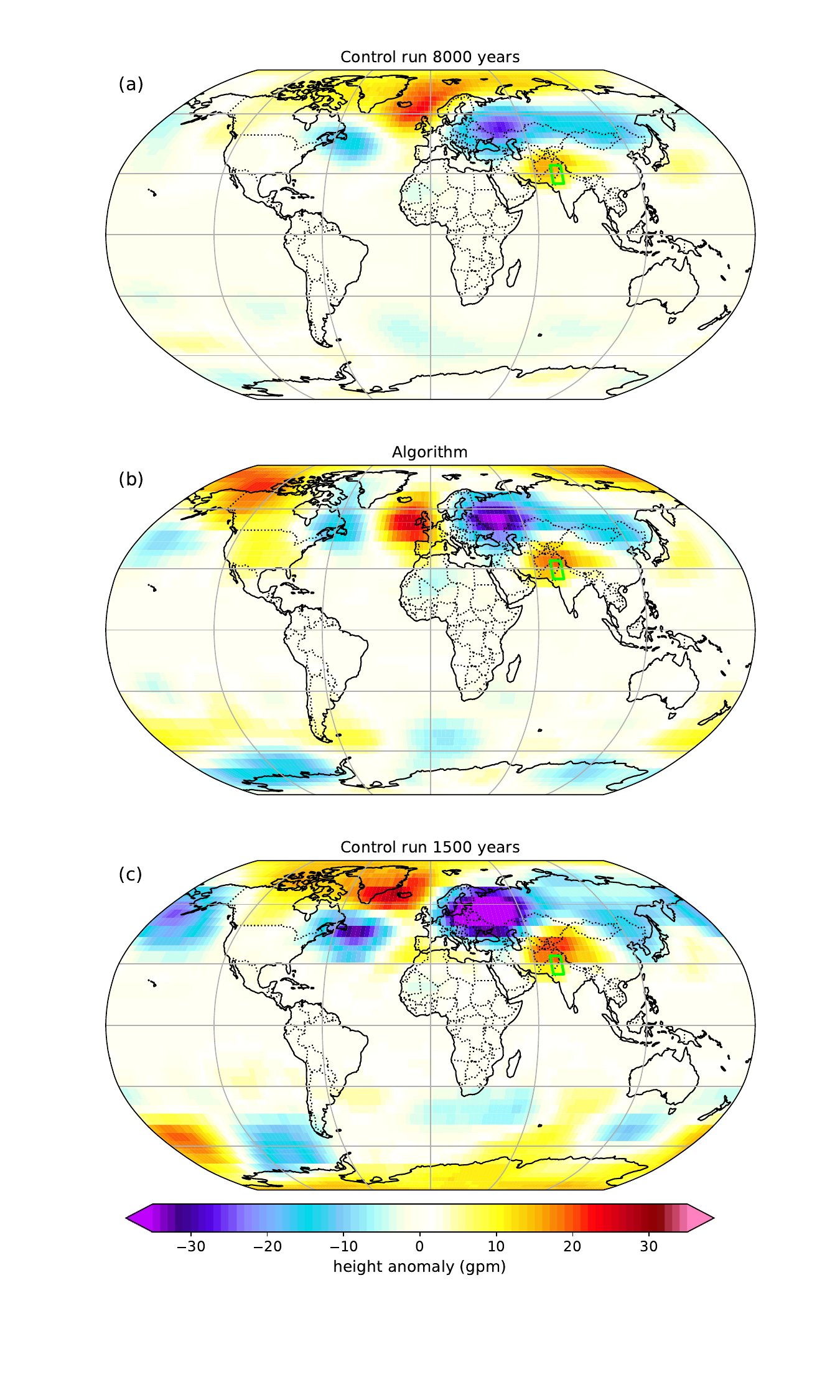} \includegraphics[width=0.49\linewidth]{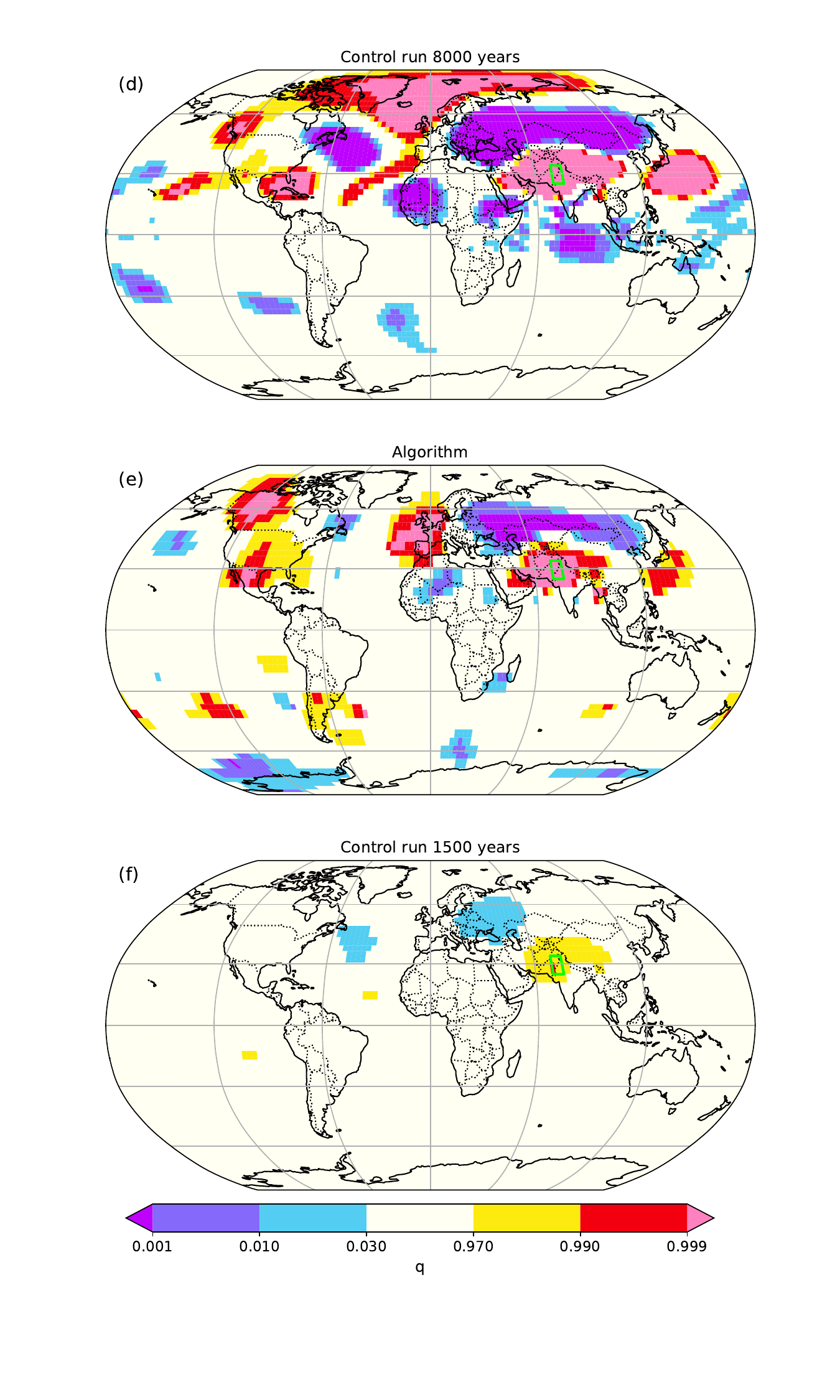}
\caption{\textit{Left:} Zg500 composite maps conditioned on $a>a_{100}$ for the 8,000 year long control run \textbf{(a)}, the rare event algorithm \textbf{(b)} and an independent control run of 1,500 years \textbf{(c)}. The maps \textbf{(a)} and \textbf{(c)} are averaged of M=80 and M=10 maps respectively while \textbf{(b)} is the mean of the 6 individual experiments. 
\textit{Right:} Corresponding significance maps. $q$ is a measure of the significance computed with a Student's t-test: the anomaly is positive (resp. negative) with probability $q$ (resp. $1-q$). See \ref{sec: Appendix confidence and significance} for details. 
The region of interest is highlighted by the green polygon. \label{fig:Zg500 100yrs HW  composites}}
\end{figure}

Rare event simulations can also be used to study dynamical quantities associated to the extreme events of interest.
Here, we study the atmospheric conditions associated to centennial heatwave seasons. 
In figure \ref{fig:Zg500 100yrs HW  composites} we compare Zg500 composite maps from the 8,000-year long control run (a) and the algorithm (b). 
The long control run composite map is an average over M=80 independent years, it is therefore an exceptionally good estimate of the ground truth for such a rare event, thanks to the unusually long dataset.
The algorithm composite map is similar to the long control run one.
In particular, the extreme heatwave seasons are associated to an anticyclonic anomaly which is centred on the northwest of the target region.
This anticyclonic anomaly is part of a large-scale hemispheric quasi-stationary wave-pattern which features anticyclonic anomalies over the North-East Atlantic, western Canada and the western Pacific ocean (East of Japan) and cyclonic anomalies at the tip of Newfoundland and over Eastern Europe. The latter anomaly stretches to the East towards the Pacific ocean.
However, the algorithm composite map presents larger anomalies in the Southern hemisphere. 
In the right column of figure \ref{fig:Zg500 100yrs HW  composites}, we plot the significance levels associated to the anomalies. 
Both the long control run and the algorithm have numerous significance regions (in particular all the anomalies mentioned above are significant). 
But the significance area for the algorithm are indeed less extended than for the 8,000-year long control run.
However, the total computation cost for running the algorithm is only 1,500 years (a 1,200-year long control run to produce the 1,200 initial conditions and 300 years of computation cost for running the algorithm itself), i.e. more than 5 times lower.
What would be the quality of a composite map computed with a control run of similar cost?
To answer this question, we add on panel (c) a composite map computed from a second independent control run of 1,500 years\footnote{In this second run there are only M=10 years in which the AMJ-averaged temperature exceeds the 100-year return level.}.
The general pattern mentioned above is still distinguishable but the map bears less resemblance with the long control run than the algorithm does (although the initial conditions for the algorithm come from this second control run.). There are also larger anomalies in the Southern ocean hinting at less significance. 
This is confirmed by the corresponding significance map (f). The 1,500-year composite map only has three regions of statistically significant anomaly. 
Therefore, it would not be possible, based on this sole map, to confidently assess the existence of the hemispheric wave-pattern mentioned above.

\begin{figure}[tbh]
\centering
\includegraphics[width=0.49\linewidth]{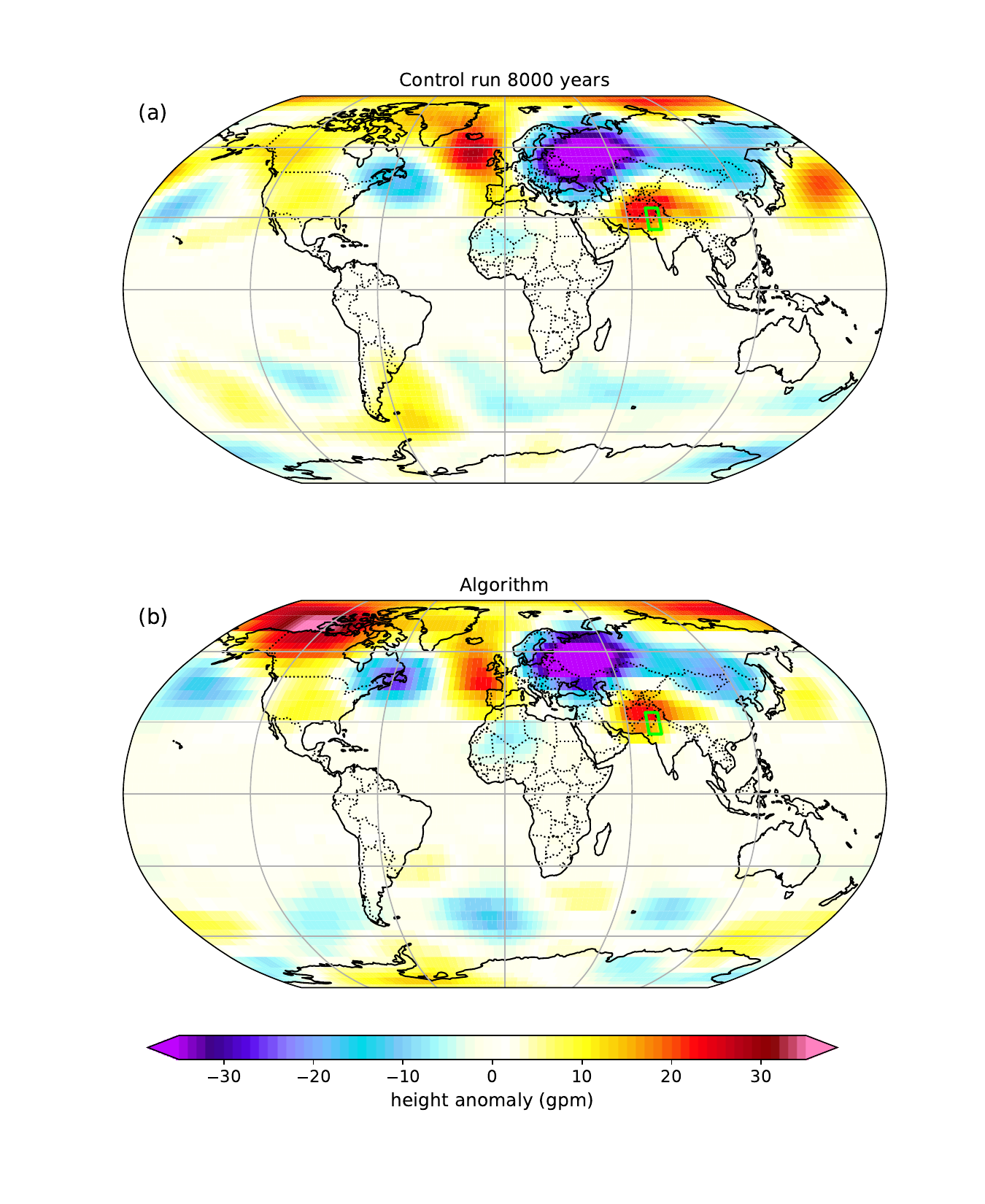} \includegraphics[width=0.49\linewidth]{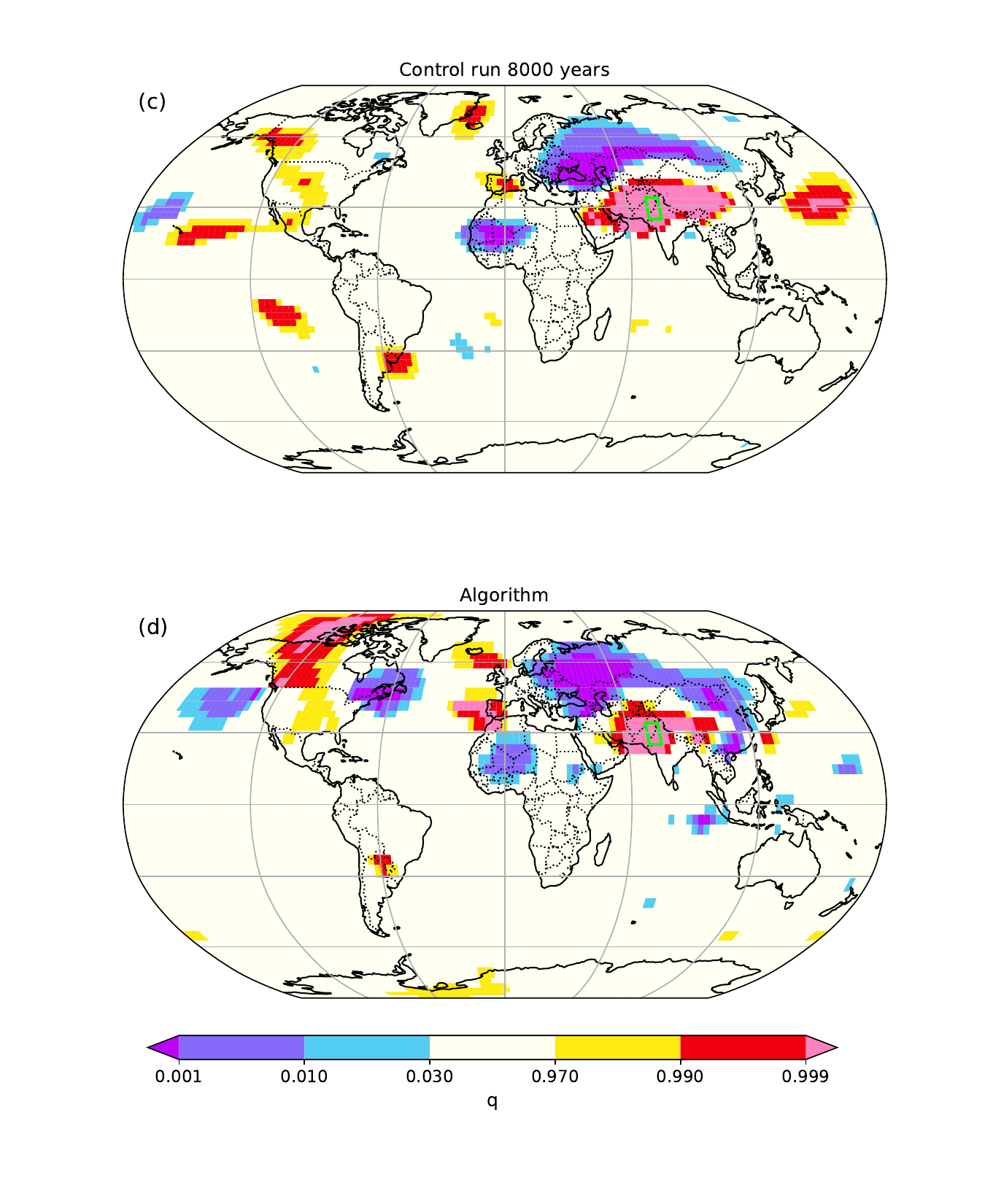}
\caption{\textit{Left:} Zg500 composite maps conditioned on $a>a_{1,000}$ for the 8,000-year long control run (average on M=8 maps) \textbf{(a)} and the rare event algorithm \textbf{(b)}.
\textit{Right:} Corresponding significance maps. The interpretation is the same as in fig. \ref{fig:Zg500 100yrs HW  composites}. 
The region of interest is highlighted by the green polygon.
\label{fig:Zg500 1,000yrs HW composites}}
\end{figure}

The algorithm is not only more efficient than a control run of equivalent cost for studying 100-year events. It can be used to study even rarer events that are out of reach of a control run at equivalent cost.
On figure \ref{fig:Zg500 1,000yrs HW composites} we show Zg500 composite maps for 1,000-year heatwave seasons.
The same threshold $a_{1,000}$, computed from the 8,000-year long control run, is used for the control run and the algorithm\footnote{There is no event exceeding the threshold in the 1,500-year long run.}.
We distinguish the same quasi-stationary wave-pattern as we observed for the centennial heatwave seasons but with stronger cyclonic and anticyclonic anomalies.  The pattern is confirmed by the significance maps (left column of fig. \ref{fig:Zg500 1,000yrs HW composites}).
Note that the significance of the algorithm is almost the same as for 100-year heatwave.
In contrast, the significance of the long control run has largely decreased.

\subsubsection{Number of centennial and millennial heatwave seasons sampled by the algorithm}

The rare event algorithm is efficient in studying centennial and millennial heatwave seasons (i.e. seasons with return times of 100 years or 1000 years) because it samples many of these seasons at a reduced computation cost. 
In this study, the algorithm sampled 1065 centennial heatwave seasons for a computation cost equivalent to 300 years. However, these seasons are not fully independent as trajectories are duplicated along the rare event simulations. Therefore it is also interesting to consider the number of independent initial conditions that yield the extreme heatwave seasons. For centennial heatwave seasons, the 1065 simulated trajectories started from 20 independent initial conditions.
Regarding 1000-year events, the algorithm simulated 527 millennial heatwave seasons originating from 12 independent initial conditions.
These numbers can be compared to the number of centennial and millennial heatwave seasons in the 8000-year long run which are (by definition) 80 and 8 respectively.
While the number of seasons is divided by 10 in the long control run when going from centennial to millennial heatwave seasons, the number of fully independent trajectories in the algorithm is only reduced by 40\%.
This shows that the relative efficiency of the algorithm improves as we shift the focus towards more extreme events.

\subsubsection{Fairness of the hemispheric wave-pattern} 

\begin{figure}[tbh]
\centering
\includegraphics[width=12cm]{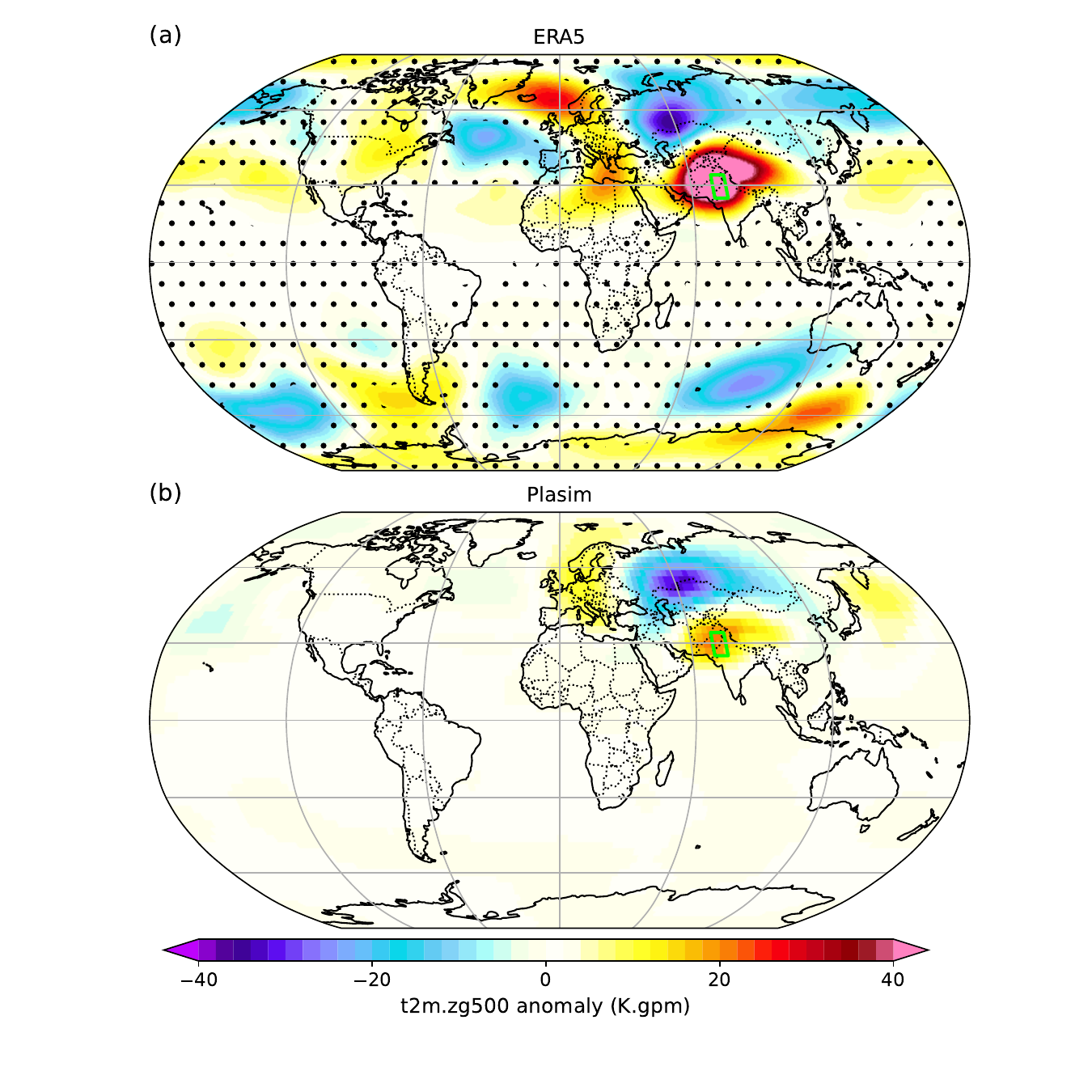}
\caption{3-day correlation maps between Zg500 and the temperature over the target region in ERA5 \textbf{(a)} and Plasim \textbf{(b)}. In \textbf{(a)}, we use a bootstrap test to assess significance at the 95\% confidence level\protect\footnotemark{}. Stippling denote regions where anomalies are non-significant. 
The region of interest is highlighted by the green polygon. \label{fig:Zg500 correlations maps Plasim ERA5}}
\end{figure}

The composite maps in figures \ref{fig:Zg500 100yrs HW  composites} and \ref{fig:Zg500 1,000yrs HW composites} show that extreme heatwave seasons in Northwest South Asia are associated to an anticyclonic anomaly centred on the northwest flank of the target region embedded in a hemispheric quasi-stationary wave-train. 
\footnotetext{The computation time to build a single correlation map in Plasim being 4 hours, it was not feasible to apply a bootstrap test in a reasonable computation time.}
Can we trust this pattern for the real world?
Of course, we cannot use historical data to build composite maps of such extreme heatwave seasons. 
So the best we can do is to test the ability of the model to represent less extreme or non-extreme statistics.
As ERA5 data-set goes back to 1940, even composite on 10-year heatwave seasons would have a poor significance. 
In order to get around this problem we assess the general ability of the model Plasim to correctly represent the teleconnections between the temperature over the target region and the 500hPa geopotential height anomaly. 
To this end we compare maps of correlation between the temperature over the target region and Zg500 computed with Plasim and ERA5 reanalysis dataset.
Before computing the correlation, we removed cubic trends from the ERA5 time series (see section \ref{sec:ERA5 data} and fig. \ref{fig:historical trends} for details) and 
coarse grain the time series by performing 3-day block average.
Figure \ref{fig:Zg500 correlations maps Plasim ERA5} show the resulting correlation maps for ERA5 (a) and Plasim\footnote{We use 1,000 years of data to compute Plasim's correlation map.} (b).
Plasim's correlation map show a strong positive correlation region centred over Pakistan and two other ones over Central Europe and the Pacific West. A wide negative correlation region centred at the border between Kazakhstan and Russia separates the positive correlations over Europe and South Asia. 
ERA5 correlation map is much noisier due to a much lower number of years (83 versus 1,000). However, the features present in Plasim's composite maps and mentioned above are all presents in ERA5 correlation map, including the negative anomaly at the tip of Newdfoundland.
This suggests that Plasim represents fairly well the teleconnection between temperature over the target region and Zg500 and that the hemispheric wave-train observed on the composite maps is something real.

\subsection{Methodological study of the algorithm}\label{sec:Methodological study}

\begin{figure}[tbh]
\centering
\includegraphics[width=1\linewidth]{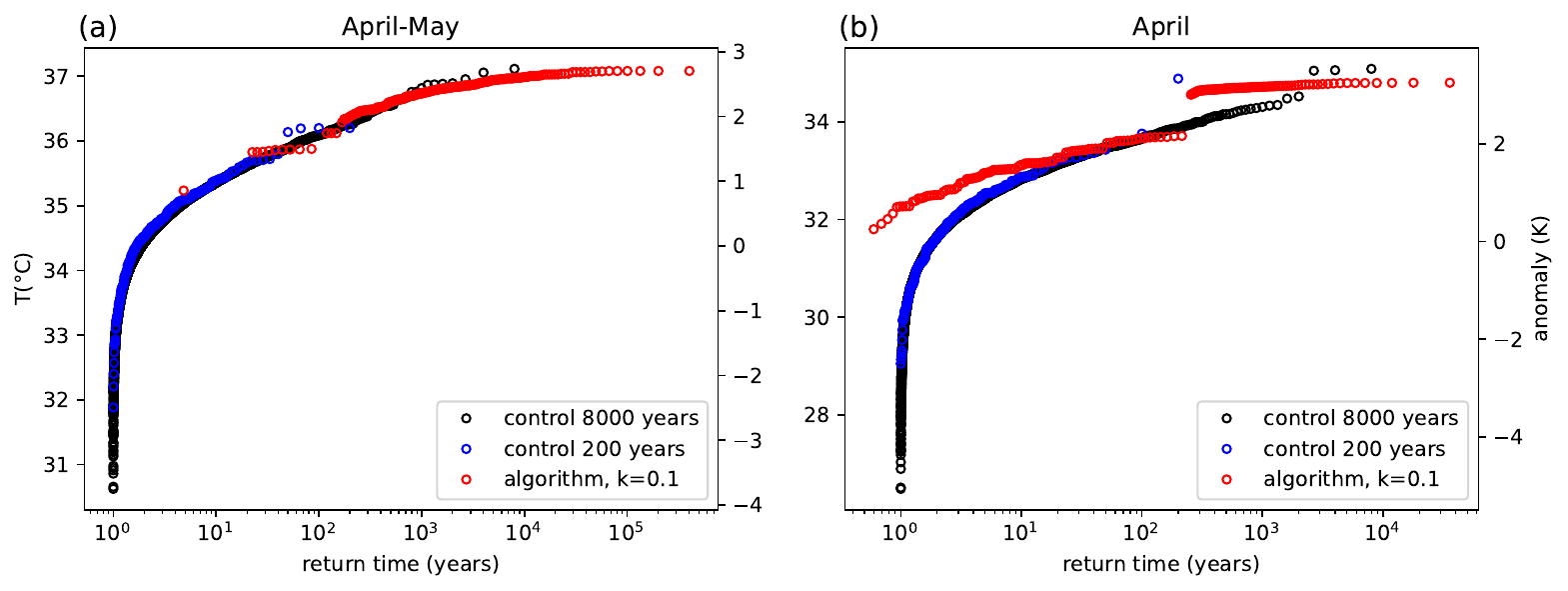}
\caption{\textbf{(a)} Return time curves of April-May averaged temperature for the long control run (black), one algorithm realisation with $N=200$ trajectories (red) and the 200-year long control run used to initialise the algorithm (blue). \textbf{(b)} Same as (a) for April averaged-temperature. \label{fig:RT curves 1&2 months averages}}
\end{figure}

Finally we address two methodological questions regarding the efficiency of the algorithm and the optimal way to run it. 
The first question pertains to the duration of extremes that the algorithm can effectively sample.
The algorithm samples extreme hot events by selecting trajectories with persistent large temperatures. 
However, it does not produce new record of the daily temperature, as can be seen on the middle line of figure \ref{fig:traj_CTRL_vs_GKTL}.
So, what is the minimal event duration for which the algorithm effectively samples extremes?
To answer this question, we show on figure \ref{fig:RT curves 1&2 months averages} the return time curves for April-May (60 days) and April (30 days) averaged temperature,
corresponding to running the algorithm for 11 and 5 resampling steps respectively\footnote{
The number of resampling steps is the event duration divided by the resampling time $\tau$ (=5 days here) minus 1 because no resampling is performed after the last integration window.}. 
For 60-day events, panel (a) shows that the algorithm effectively samples extremes up to 1K hotter than in the control run used for initialisation. Moreover, the algorithm curve is in good agreement with the long control run.
For 30-day events, panel (b) shows that the algorithm does not sample more extreme events than what is already present in the initialisation control run. Many algorithm trajectories stem from the same initial condition which corresponds to an outlier April temperature of the initialisation run. The remaining trajectories do not exceed the second largest April temperature of the initialisation run. Finally, the algorithm curve falls off the control run curve at low return time, with trajectory probabilities being overestimated.

While this result suggests that the algorithm does not improve the sampling of extreme 30-day events, we must mention that it is actually the case if one follows a slightly different approach than ours: when considering the largest 30-day averaged anomaly during a four month long period, the algorithm does greatly improve the sampling of extreme 30-day events (see chapter 5 of the PhD thesis \cite{cozian2023}).

A possible alternative way to sample shortest events could be to decrease the resampling time $\tau$ while keeping the number of resampling steps constant. Although we did not try out this approach, we do not expect it to work.
Indeed, an analysis of the trajectories reveals that clone trajectories barely separate after 5 days. 
If $\tau$ were smaller, clone trajectories would not separate during the integration window and the subsequent resampling would not discriminate among trajectories.

\begin{figure}[tbh]
\centering
\includegraphics[width=1\linewidth]{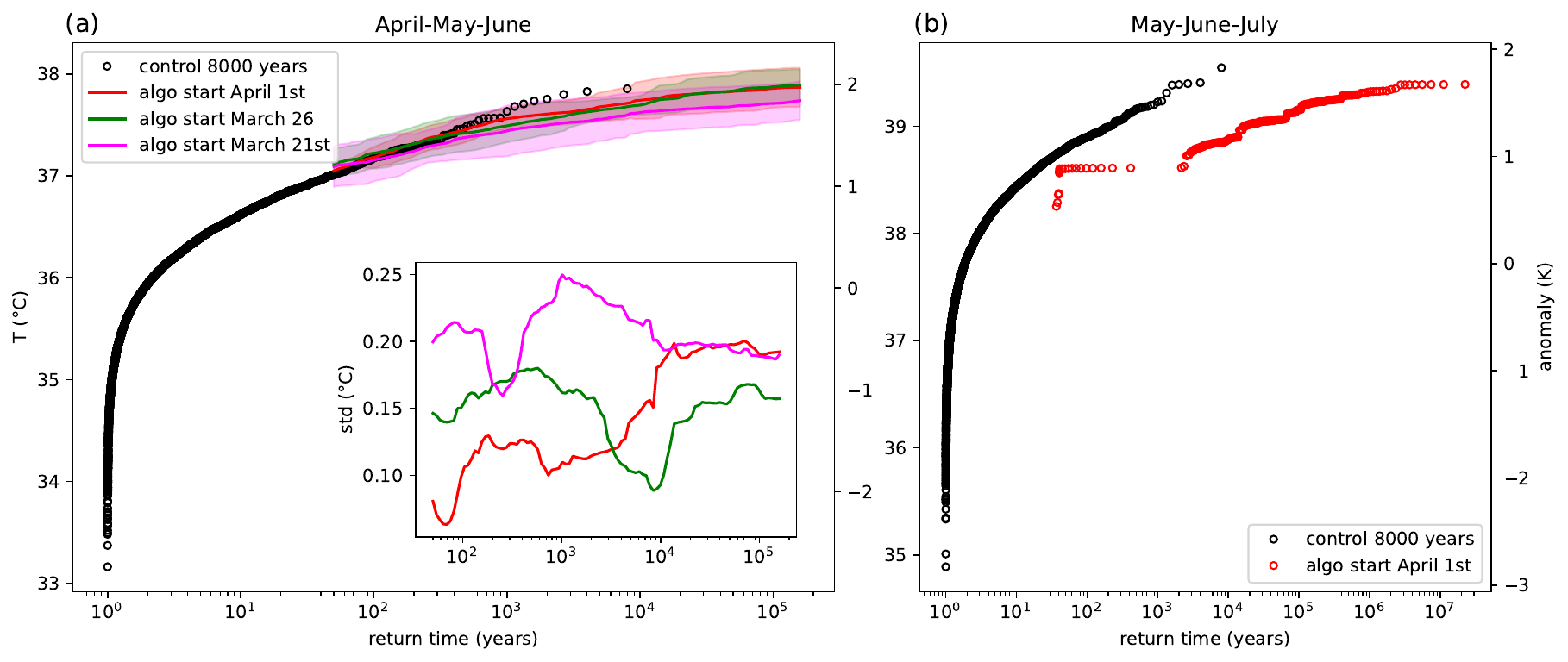}
\caption{\textbf{(a)} Return time curves of April-May-June averaged temperature for the long control run (black) and three sets of algorithm experiments with starting dates on April 1st (red), March 26 (green) and March 21st (magenta). M=6 experiments were run for each starting date. Solid lines and shaded areas represent the inter-experiments mean and standard deviation respectively.
\textbf{Inset:} Inter-experiments standard deviation for the three different starting dates.
\textbf{(b)} Return time curves of May-June-July averaged temperature for the long control run (black) and one algorithm realisation that was started on April 1st. \label{fig:RT curves earlier start}}
\end{figure}

The second question is whether the algorithm is more effective when started a few resampling times ahead of the period studied.
The underlying idea is that the algorithm could need a few resampling steps before reaching a statistical equilibrium among trajectories.
On panel (a) of figure \ref{fig:RT curves earlier start} we compare the set of algorithm experiments starting at the beginning of the period studied (April 1st for AMJ averaged temperature) with two other sets of experiments starting one and two resampling times earlier respectively (i.e. 5 and 10 days earlier). 
Each set contains $M=6$ algorithm experiments with $N=200$ trajectories. 
The return time curve of the set starting 5 days earlier is barely distinguishable from the standard experiment set.
The uncertainty is smaller or larger than for the standard experiment depending on the return time. Thus, no gain is achieved globally on the precision of the estimated return levels. 
The 10-day earlier start curve is slightly under the two others but remains within the confidence intervals. Its inter-experiment standard deviation is consistently larger, thus yielding a larger uncertainty on the estimates. 
How to discriminate between the different curves?
Is the lower curve of the 10-day earlier start a better estimate or does it suffer from a systematic bias?
Panel (b) of figure \ref{fig:RT curves earlier start} shows the effect of a start of the algorithm 30 days ahead of the period of interest. We show the return time curves for MJJ averaged temperature with the algorithm being started on April 1st.
The algorithm falls completely off the control run curve by largely underestimating the events probabilities.
The reason is that large temperatures in April give high weights, thus low probabilities, to the trajectories but do not enter the considered temperature average. 
It is probable that the 10-day earlier start suffers from the same bias, although in a smaller way.
We conclude that an earlier start does not improve the efficiency of the algorithm and can actually be detrimental to the quality of the results by inducing a systematic bias where probabilities are underestimated.

\section{Relevance for extreme event attribution}\label{sec:relevance of EEA}

We showed that a rare event algorithm can provide better estimates of return levels than an EVT fit, with confidence intervals that are much narrower for large return times.
Therefore, such algorithms could be used in attribution studies to better constrain the change in return time - or return level - with global warming.
It would require running the algorithm at least two times, one for the factual world and one for the counter-factual world. Additional runs could be required if one wants to explore the change in a warmer (e.g. +2°C) world.
However, running the algorithm requires much more work (and computation hours) than performing a statistical fit on already existing data. 
Therefore one needs to consider if it is worth the effort.
We give here a few considerations on this question.

We showed that the gain in precision on return levels increases drastically with the rarity of the events. Therefore rare event algorithms would be particularly suited for the rarest events.
It demands probably too much work for rapid attribution studies but could be used to refine the estimates in subsequent peer-reviewed studies.
Robust attributions are performed when many models are combined with observations to obtain a \textit{hazard synthesis} \cite{philip2020}. 
When there are many valid models, one might not want to run the algorithm for each model but it could be used to refine the estimate in the model with the highest uncertainties. 
In the alternative when there are only few valid models, one could envision to run it for all models.

We want to emphasise that an important added-value of rare event algorithms is that they actually sample the extreme events of interest, thereby allowing to compute composite statistics with good significance. Therefore they could be used for process-based studies and investigate how 
the process associated to very extreme events change with global warming.

Our method bears some resemblance to the UNprecedented Simulated Extreme ENsemble (UNSEEN) methodology \cite{thompson2019, kelder2020}. UNSEEN makes use of past ensemble seasonal forecasts or hindcasts that provide several hundreds years long datasets of plausible meteorological conditions. The large datasets of ensemble seasonal forecasts provide many events that exceed observational records and allow to draw empirical return time curves up to about a millenia that can be prolonged with EVT fits \cite{vandenbrink2004, kelder2020}. UNSEEN has been used, for instance to reduce the confidence interval on $10^4$-year surge level estimates in the Netherlands \cite{vandenbrink2004} or to detect and quantify changes in 100-year precipitation extremes over the West coast of Norway in the last decades \cite{kelder2020}.
Since it make use of archives of operational weather forecasts, UNSEEN does not require to run dedicated model experiments. 
However, rare event algorithms allow to sample more extremely rare events than UNSEEN does. For instance we were able to sample many millennial heatwave seasons.

Rare event algorithms and UNSEEN could be combined together: the initial conditions required to start the algorithm could be drawn from the ensemble seasonal forecasts simulations. 
Note that the algorithm can be applied to any model, with a change in the perturbation that must be adapted to each model. It has already been applied to a state of the art GCM \cite{ragone2021}.
Such a coupling of the rare event algorithm with ensemble seasonal forecasts would allow to sample rarer extreme events for a marginal additional computation cost. 
It represents a promising avenue for the application of rare event algorithms to extreme event attribution.

\section{Conclusion}\label{sec:Conclusion}

In this study, we demonstrate how a rare event algorithm samples extremely rare events in a climate model and allows the computation of various related statistics.
We compared the performance of the algorithm against a 8,000-year long control run which enabled us to demonstrate the accuracy of the statistics computed via the algorithm.
First, we showed that the algorithm provides a good estimate of the return time curve with narrow confidence intervals.
Secondly, since it actually samples extreme events, the algorithm can be used to compute statistics conditioned on the extreme event of interest. 
For instance, composite maps conditioned on 100-year heatwave seasons were found to be in good agreement with the long control run estimates.
These maps suggest that extreme heatwave seasons in South Asia are associated with an anticyclonic anomaly which is embedded in a large-scale hemispheric quasi-stationary wave-pattern.
Comparison with reanalysis give confidence in this result, since the model represents well teleconnections between the 2-meter temperature over the target region and Zg500 over the Northern hemisphere. 

The algorithm also successfully estimates the intensity-duration-frequency statistics of individual subseasonal heatwaves. Thus, the algorithm can also be used to estimate the distribution of intensity and duration of heatwaves that comprise a heatwave season.
Thus rare event algorithms could, for instance, be combined with seasonal forecasts to provide information regarding expected number of heatwave days and the distribution of the duration and intensity in an extreme heatwave season. This combination of seasonal mean and subseasonal variability could be useful for adaptation planning or risk management. In addition, the algorithm can be used to generate more samples of individual heatwaves and heatwave seasons.
Such samples could be used in analysis of circulation features relevant to the predictability of seasonal weather such as atmospheric waveguides \cite{white2022}, for process-oriented studies and data-driven seasonal forecasting methods.

Note that the efficiency of the algorithm does not depend on the choice of the region, nor on its size.
If we were to consider a different region, region size or averaging period, the variability of the target observable would change and thus the selection parameter $k$ should be adapted in consequence. But apart from adjusting this parameter, the algorithm would be as effective. 
It can as well be applied at the grid point scale for a season or to the whole globe over a year to sample anomalously warm years.
In fact, the algorithm has already been applied at a continental scale \cite{ragone2018} and on multi-year time scales \cite{ragone2020}.

Most current literature on use of rare event algorithms to climate models, including our work, have analysed heatwaves as measured by air temperature, and in dynamical regimes where Rossby waves are the primary drivers of such heatwaves. 
Application to humid heatwaves (and other extreme events of interest such as droughts or precipitations) and to tropical regions remains an avenue for future research. Regardless, rare event algorithms are a promising methodological advance which allows for a systematic study of the statistics and dynamics of extreme events in climate models.

\begin{ack}
This project was provided with computing and storage resources by GENCI at TGCC on the partition ROME of supercomputer Joliot Curie thanks to the grant 20XX-A010110575.
Data analysis was carried out using the resources of the Centre Blaise Pascal at ENS de Lyon. We are grateful to Emmanuel Quemener for his help with the platform.
This work was supported by the ANR grant SAMPRACE, Project No. ANR-20-CE01-0008-01 (C.L.P).
C.L.P. thanks F. Ragone for sharing its code of the algorithm, B. Cozian for help with running Plasim and D. Faranda for advices on EVT fits.
\end{ack}

\section*{Data availability statement}
The scripts and processed data used for this study are available on the following Zenodo repository: \href{https://doi.org/10.5281/zenodo.10888194}{https://doi.org/10.5281/zenodo.10888194}.
This repository contains an example script to run the algorithm, sampling outputs from the algorithm, and the scripts used to perform the analysis and produce the figures in the present article.

\newpage

\appendix

\section{Detailed description of the algorithm}\label{sec:Details of the algorithm}

\paragraph{Estimation of the biasing parameter $k$}
When running the algorithm, one must make a choice for the value of the biasing parameter $k$. This choice must be adapted to the goal of the study, more specifically to the target value of temperature anomaly $a$. To avoid a trial and error approach, we need to estimate a priori the order of magnitude $k$. We give here a heuristic derivation of this estimate by dimensional analysis. A more rigorous derivation, using large-deviation theory can be found in \cite{ragone2020}.
First, note from \eref{eq:weights formula} that $[k] = [A]^{-1} [t]^{-1}$ where $[..]$ denotes the dimension of variables.
Note also that $k$ changes sign under the transformation $a \rightarrow -a$.
Clearly, $|k|$ must increase with $|a|$ and $k$ changes sign under the transformation $a \rightarrow -a$.
Moreover, it should decrease with the spread of the distribution 
$\frac{1}{T_a}\int_{t_0}^{t_0+T_a} A(t) dt$ (as it is easier to reach a given level if the variance is larger).
Let $\Sigma$ be the standard deviation of the latter distribution.
Introducing $T_a$ as a characteristic time of the algorithm, we can write 
$k=a^{\alpha}  \Sigma^{\beta} T_a^{\gamma}$.
The dimensional analysis imposes that $\alpha+\beta=-1$ and $\gamma=-1$. For the first relation, any solution with $\alpha>0$ and $\beta<0$
would meet the requirement. 
However, the anti-symmetry w.r.t. $a \rightarrow -a$ implies that $\alpha$ is an odd integer.
We stick to the simplest solution $\alpha=1$, $\beta=-2$. 
Our estimate of the optimal $k$ is thus:
\begin{equation}
    k^* = \frac{a}{T_a \Sigma^2}.
\end{equation}
Note that the greater $k$, the more stringent the selection is. If $k$ is too large compared to $N$, it can be the case that only a single initial trajectory survives after a few selection steps. We are thus loosing the diversity of the initial condition and the result can be largely biased. To avoid this issue, $N$ must be large enough so that diversity of trajectories is preserved.

\paragraph{Computation of the probabilities}
Trajectories that survive until the end of the algorithm are reconstructed by attaching to trajectories at step $i+1$ their parent trajectories at step $i$, starting from the last resampling step of the algorithm.
At each time step of the algorithm the probability of each trajectory to survive is biased by a factor $W_n^{(i)}$.
We thus obtain a biased distribution
$\mathbb{P}_k \left( \left\lbrace x(t) \right\rbrace_{0\leq t \leq T_a} \right)$
over the trajectories which is linked to the unbiased probability 
$\mathbb{P}_0\left( \left\lbrace x(t) \right\rbrace_{0\leq t \leq T_a} \right)$ 
to observe the trajectories under the model dynamics by the relation \cite{ragone2018}:
\begin{eqnarray}\label{eq: relation P_k P_0}
\mathbb{P}_k
\left( \left\lbrace x_n(t) \right\rbrace_{0\leq t \leq T_a} \right)
&= \left( \prod_{i=1}^{I-1} W_n^{(i)} \right) \mathbb{P}_0 \left( \left\lbrace x_n(t) \right\rbrace_{0\leq t \leq T_a} \right) \\
 &= \frac{\exp \left( k\int_0^{T_a} A_n(t)dt \right)}{\tilde{Z}}  \mathbb{P}_0 \left( \left\lbrace x_n(t) \right\rbrace_{0\leq t \leq T_a} \right).
\end{eqnarray}
$\tilde{Z}$ is a normalisation factor that can be computed from the algorithm as $\tilde{Z} = \prod_{i=1}^{I-1} Z_i$.
By inverting equation \eref{eq: relation P_k P_0} we can obtain an unbiased estimator for any observable 
$O\left( \left\lbrace x(t) \right\rbrace_{0\leq t \leq T_a} \right)$
of the dynamics:
\begin{eqnarray}
\mathbb{E}_0\left[ O\left( \left\lbrace x(t) \right\rbrace_{0\leq t \leq T_a} \right) \right] \sim_{N\to+\infty}
\sum_{n=1}^N p_n O\left( \left\lbrace x_n(t) \right\rbrace_{0\leq t \leq T_a} \right) \label{eq:empirical observable estimator} \\
\textrm{with } p_n = \frac{1}{N} \left( \prod_{i=1}^{I-1} W_n^{(i)} \right)^{-1} = \frac{1}{N}\frac{\exp \left( k\int_0^{T_a} A(t)dt \right)}{\tilde{Z}} \label{eq:formula p_n}
\end{eqnarray}
The probabilities $p_n$ can be readily computed from the algorithm by keeping track of the weights $W_n^{(i)}$.
Formulae (\ref{eq:empirical observable estimator}, \ref{eq:formula p_n}) allow us to compute empirical expectations of any observable by performing an ensemble average weighted by the probabilities $p_n$.
When we have several experiment, expectations are computed separately for each independent experiment. Then we perform a second average over the experiments.

Note that, in practice, the rate of convergence in \eref{eq:empirical observable estimator} depends on the observable $O$. 
The rate of convergence is good (efficiency of the algorithm) to compute
$\mathbb{P}_0 \left( \frac{1}{T_a}\int_0^{T_a} A(t) dt > a \right) = \mathbb{E}_0\left[ \mathbf{1}\left( \frac{1}{T_a}\int_0^{T_a} A(t) dt > a \right) \right]  $
and composite statistics conditioned on $\frac{1}{T_a}\int_0^{T_a} A(t) dt > a$,
provided that  we choose a biasing parameter $k$ adapted to the target value $a$.
It would however be slower than a classical ensemble simulation to compute unconditioned statistics.

\section{Estimation of confidence intervals and significance levels}\label{sec: Appendix confidence and significance}

\paragraph{Confidence intervals}
Let $\{X_i \}_{1\leq i \leq n}$ be $n$ i.i.d. samples from a random variable $X$.
We want to build a confidence interval for the true mean $\mu_X$ of $X$.
We compute its empirical mean $\bar{X} = (\sum_{i=1}^n X_i ) /n$ and unbiased standard deviation estimator 
$S = \sqrt{ \sum_{i=1}^n (X_i - \bar{X})^2 /(n-1) }$.
We assume that the random variable
\begin{equation}
    T:=  \frac{\bar{X} - \mu_X}{S / \sqrt{n}}
\end{equation}
follows the $t$-distribution with $(n-1)$ degrees of freedom $t_{n-1}$.
We note $t_{\nu, \alpha}$ the quantile of order $1-\alpha$ from the distribution $t_{\nu}$.
Then $T$ belongs to the interval $[-t_{n-1, \alpha/2}, t_{n-1, \alpha/2} ]$ with probability $1-\alpha$. This is equivalent to:
\begin{eqnarray}\label{eq: t-test confidence interval}
    \mathbb{P} \left( \mu_X \in \left[ \bar{X} - t_{n-1, \alpha/2}\frac{S}{\sqrt{n}} ; \bar{X} + t_{n-1, \alpha/2}\frac{S}{\sqrt{n}} \right] \right) = 1-\alpha .
\end{eqnarray}
We use \eref{eq: t-test confidence interval} with $\alpha=0.05$ to build 95\% confidence intervals. 

\paragraph{Significance levels}
We compute $\tilde{T}:=  \sqrt{n} \bar{X} / S$ 
and $q=F_{n-1} (\tilde{T})$ where $F_{n-1}$ is the cumulative distribution function of $t_{n-1}$. 
$\tilde{T}$ is a test of the null hypothesis that $\mu_X=0$.
If $q > 1/2$ then $\mu_X > 0$ with probability $q$. Conversely if $q< 1/2$ then $\mu_X < 0$ with probability $1-q$

\section{Extreme Value Theory fit}\label{sec: EVT fit details}

\begin{figure}[htb]
\centering
\includegraphics[width=0.6\linewidth]{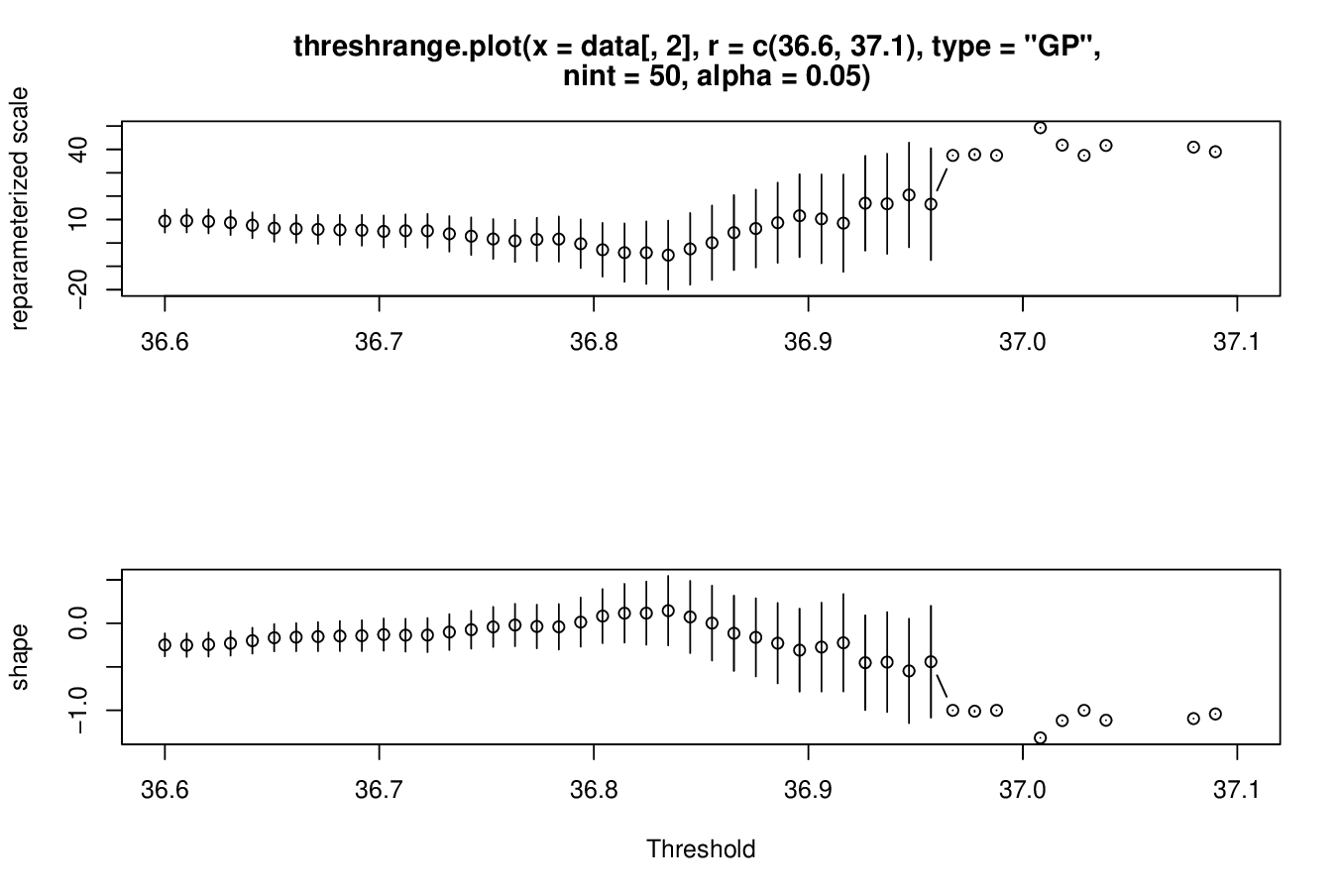} 
\caption{Reparametrized scale and shape parameters estimates for different choice of the threshold. \label{fig:threshrange_plot}}
\end{figure}

\begin{figure}[htb]
\centering
\includegraphics[width=0.8\linewidth]{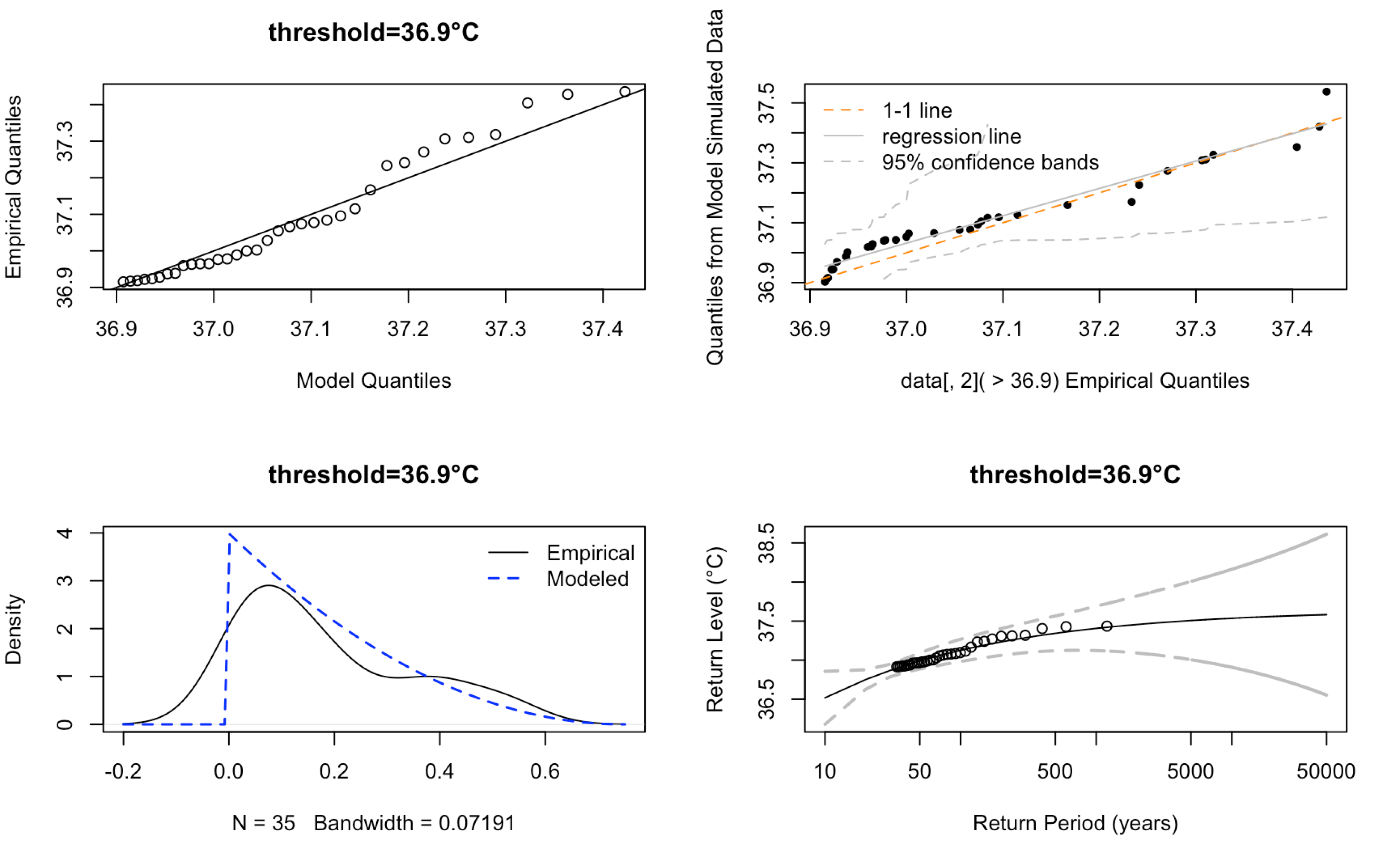}
\caption{Diagnostic plots for the POT fit with threshold 36.9°C . 
Quantile-quantile plot (top left), quantiles from a sample drawn from the fitted GPD distribution against the empirical data quantiles with 95\% confidence bands (top right), density plots of empirical data and fitted GPD distribution (bottom left), and return level plot with 95\% pointwise normal approximation confidence intervals (bottom right). \label{fig:EVT_fit_summary}}
\end{figure}

We performed an EVT fit, using the Peak Over Threshold approach with a threshold of 36.9°C. We use R package \textit{extRemes 2.0} \cite{gilleland2016}.
Figure \ref{fig:threshrange_plot} shows the estimates for different range of the threshold, supporting the choice of 36.9°C (see sections 4.3.4 and 4.4 of \cite{coles2001} for details).
The resulting estimate for the shape parameter is $\xi=-0.336$ ($[-0.470, 0.235]$ 95\% C.I.) and the scale parameter is $\sigma=0.251$  ($[0.179, 0.440]$ 95\% C.I.).
Figure \ref{fig:EVT_fit_summary} shows diagnostic plots of the estimated model.
The confidence intervals on the fit parameters and return levels are estimated via the profile likelihood method.

In figure \ref{fig:PDF & RT curves}
we compare the 95\% confidence interval of the algorithm with the one from the EVT fit estimated via two different methods: profile likelihood and a non-parametric bootstrap.
We give below a brief description of each method.

\paragraph{Profile likelihood} 
Suppose the EVT model is determined by two parameters $(\theta^{(1)}, \theta^{(2)})$. 
Let $\ell(\theta^{(1)}, \theta^{(2)})$ be the log-likelihood of the model. 
The profile likelihood along $\theta^{(1)}$ is defined as
$\ell_p(\theta^{(1)}) = \max_{\theta^{(2)}} \ell(\theta^{(1)}, \theta^{(2)})$.
This can be generalized to an arbitratry number of parameters.
Let $\hat{\theta}_0$ be the maximum likelihood estimate of the model $(\theta^{(1)}, \theta^{(2)})$.
Under suitable regularity conditions:
\begin{equation}
    D_p(\theta^{(1)}) = 2 \left( \ell(\hat{\theta}_0) - \ell_p(\theta^{(1)}) \right) \sim \chi_1^2 .
\end{equation}
Let $c_{\alpha}$ be the $(1-\alpha)$ quantile of the $\chi_1^2$ distribution. 
Then $C_{\alpha} = \{\theta^{(1)} | D_p(\theta^{(1)}) \leq c_{\alpha} \} $ is a $(1-\alpha)$ confidence interval for $\theta^{(1)}$.
A more comprehensive explanation can be found in sections 2.6.6 and 2.7 of \cite{coles2001}.
The method is implemented in R \textit{extRemes 2.0}.

\paragraph{Non-parametric bootstrap}
The generalized Pareto distribution (GPD) was fitted on the exceedances over the 36.9°C threshold.
We generated 2000 bootstrap samples by resampling with replacement from the observed exceedances and fitted a GPD for each sample.
For each return time, we define the lower (resp. higher) bound of the 95\% CI as the 2.5 (resp. 97.5) quantile of the bootstrap distribution.

\clearpage

\section*{References}
\bibliographystyle{apalike}
\bibliography{Extreme_HW_seasons_v2.bib}

\end{document}